\documentclass[12pt,thmsa]{article}
\usepackage{amssymb}
\usepackage{amsmath}
\usepackage{graphicx}
\newcommand{\esssup}{\operatornamewithlimits{ess\sup}}

\def\R{\mathbb{R}}
\def\C{\mathbb{C}}
\def\N{\mathbb{N}}

\newtheorem{theorem}{Theorem}
\newtheorem{corollary}[theorem]{Corollary}
\newtheorem{lemma}[theorem]{Lemma}
\newtheorem{proposition}[theorem]{Proposition}
\newtheorem{definition}[theorem]{Definition}

\newtheorem{remark}[theorem]{Remark}

\begin{document}

\author{\textbf{Yuri G.~Kondratiev} \\
{\small Fakult\"at f\"ur Mathematik, Universit\"at Bielefeld,
D 33615 Bielefeld, Germany}\\
{\small Forschungszentrum BiBoS, Universit\"at Bielefeld, 
D 33615 Bielefeld, Germany}\\
{\small Germany-Ukraine Center of Interdisciplinary Research and Education,}\\ 
{\small National University of ``Kyiv-Mohyla Academy'', Kiev, Ukraine}\\
{\small Laboratory of Complex Systems, NPU, Kiev, Ukraine}\\
{\small kondrat@mathematik.uni-bielefeld.de} \and 
\textbf{Tobias Kuna} \\
{\small Center for Mathematical Sciences Research,}\\ 
{\small Rutgers University, New Jersey, USA}\\
{\small Forschungszentrum BiBoS, Universit\"at Bielefeld,
D 33615 Bielefeld, Germany}\\
{\small tkuna@mathematik.uni-bielefeld.de} \and
\textbf{Maria Jo\~{a}o Oliveira} \\
{\small Universidade Aberta, P 1269-001 Lisbon, Portugal}\\
{\small Centro de Matem\'atica e Aplica{\c c}\~oes Fundamentais,}\\ 
{\small University of Lisbon, P 1649-003 Lisbon, Portugal}\\
{\small Forschungszentrum BiBoS, Universit\"at Bielefeld,
D 33615 Bielefeld, Germany}\\
{\small oliveira@cii.fc.ul.pt}}

\title{Non-equilibrium stochastic dynamics of continuous systems and 
Bogoliubov generating functionals}
\date{}
\maketitle

\vspace*{-1cm}

\begin{abstract}
Combinatorial harmonic analysis techniques are used to develop new 
functional analysis methods based on Bogoliubov functionals. Concrete 
applications of the methods are presented, namely the study of a 
non-equilibrium stochastic dynamics of continuous systems. 
\end{abstract}



\section{Introduction}

The combinatorial harmonic analysis on configuration 
spaces introduced and developed in \cite{KoKu99}, \cite{KoKu99c}, \cite{K00}, 
\cite{KoKuOl00b} is a natural tool for the study of equilibrium states of 
continuous systems in terms of the corresponding Bogoliubov or generating 
functionals. Originally, this class of functionals was introduced by 
N.~N.~Bogoliubov in \cite{Bog46} to define correlation functions for 
statistical mechanics systems. In 
the context of classical statistical mechanics, this class of functionals, as 
a basic concept, was analyzed by G.~I.~Nazin. We refer to \cite{Naz85} for 
historical remarks and references therein. Apart from this specific 
application, and many others, the Bogoliubov functionals are, by 
themselves, a subject of interest in infinite dimensional analysis. This is 
partially due to the fact that to any probability measure $\mu$ defined on the 
space $\Gamma$ of locally finite configurations one may associate a 
Bogoliubov functional
$$
L_\mu(\theta) :=\int_\Gamma \prod_{x\in \gamma }(1+\theta(x))\,d\mu(\gamma),
$$
allowing the study of $\mu$ through the functional $L_\mu$. 
Technically, this means that through the Bogoliubov functionals one may 
reduce measure theory problems to functional analysis ones, yielding a new 
method in measure theory as well as new applications in functional analysis. 

From this standpoint, new perspectives were announced in \cite{KoKuOl00b} 
in the setting of combinatorial harmonic analysis on configuration spaces. The 
purpose of this work is to carry out these technical improvements. 

Of course the domain of a Bogoliubov functional $L_\mu$ depends on the 
underlying probability measure $\mu$. Conversely, the domain of a Bogoliubov 
functional $L_\mu$ carries special properties over to the probability measure 
$\mu$. In this work we mainly analyze the class of entire Bogoliubov 
functionals on a $L^1$-space (Section \ref{Section3}), which is a natural 
environment to widen the scope of this work towards Gibbs measures (or 
equilibrium states). This restriction allows, in particular, to recover the 
notion of correlation function. 

As a side remark, let us mention that in the same setting further 
progresses, under analytical assumptions, are achieved in \cite{Ku05} on a 
space of continuous functions.  

The close relation between probability measures and Bogoliubov functionals 
is best illustrated by a ``dictionary'' (cf.~G.~I.~Nazin), relating 
measure concepts and problems to functional analysis ones. In this 
``dictionary'', the translation of the Dobrushin-Lanford-Ruelle equation, 
defining Gibbs measures, leads to a functional equation, called the 
Bogoliubov (equilibrium) equation (Section \ref{Section4}). As a result, 
through analytical techniques one may derive a uniqueness result for 
Gibbs measures corresponding to positive potentials in the high 
temperature-low activity regime (Theorem \ref{9Prop9.2.8}). Although this 
result does not improve the known uniqueness results for Gibbs measures (see 
e.g.~\cite{DSI75}, \cite{PZ99}, \cite{Ru63}), its proof is technically new and 
presents an alternative approach to the uniqueness problem. 

This work concludes with a concrete application of the Bogoliubov functionals 
to the study of a non-equilibrium diffusion dynamics of a continuous system 
(Section \ref{Section5}). For particles in suspension in a liquid, each 
particle interacts with the molecules of the fluid and the remaining particles 
in the suspension. At the microscopic level, the time evolution of the whole 
system is described by Hamiltonian dynamics. In the mesoscopic 
approximation, the system is described as the result of random perturbations 
of the particles with dynamics heuristically given by a system of 
stochastic differential equations 
\begin{equation}
\left\{ 
\begin{array}{l}
\displaystyle dx_k(t)=-\frac\beta 2\sum_{1\leq i\neq k}
\nabla V(x_k(t)-x_i(t))dt+dW_k(t),\quad t\geq 0 \\ 
\\ 
x_k(0)=x_k,\quad k\in \N
\end{array}
\right. \label{9Eq3.5}
\end{equation}
for a given starting configuration $\gamma =\{x_k:k\in\N\}$. Here $W_k$, 
$k\in\N$, is a family of independent standard Brownian motions
describing the random perturbations and $V:\R^d\backslash\{0\}\rightarrow\R$ 
is the interaction potential between the particles. The problem of existence 
of a stochastic dynamics corresponding to the system (\ref{9Eq3.5}) has been 
well analyzed for the equilibrium stochastic dynamics case (see 
e.g.~\cite{AKR97a}, \cite{O96}, \cite{Y96}). For non-equilibrium dynamics, 
the existence problem is essentially open and at the moment
all we have is the construction of non-equilibrium processes done in
\cite{Fr87}, in the case of smooth potentials with finite range and $d\leq 4$, or the existence of time evolution for correlation functions described by a
correlation diffusion hierarchy (see \cite{KRR02} and the references
therein). Our goal now is the study of the non-equilibrium case in terms of 
Bogoliubov functionals. The procedure that is used turns out to be an 
effective method for the study of other equilibrium and non-equilibrium 
problems for continuous systems. Further examples of applications, e.g., 
equations for birth-and-death and hopping type dynamics on configuration 
spaces in terms 
of Bogoliubov functionals, are now being studied and will be reported in 
forthcoming publications. 

\section{Harmonic analysis on configuration spaces\label{Section2}}

Let $X$ be a geodesically complete connected oriented non-compact Riemannian 
$C^\infty $-manifold and $\Gamma:=\Gamma_X$ the configuration space over $X$:
$$
\Gamma :=\left\{ \gamma \subset X:\left| \gamma\cap K \right| <\infty 
\hbox{
for every compact }K\subset X\right\} .
$$
Here $\vert \cdot \vert$ denotes the cardinality of a set. As usual we 
identify each $\gamma \in \Gamma$ with the non-negative Radon measure 
$\sum_{x\in \gamma}\varepsilon_x\in \mathcal{M} (X)$, where 
$\varepsilon_x$ is the Dirac measure with mass at $x$, 
$\sum_{x\in \emptyset}\varepsilon_x$ is, by definition, the zero measure, and 
$\mathcal{M} (X)$ denotes the space of all non-negative Radon measures on 
the Borel $\sigma$-algebra $\mathcal{B}(X)$. This procedure allows to 
endow $\Gamma$ with the topology induced by the vague topology on 
$\mathcal{M} (X)$. We denote the Borel $\sigma$-algebra on $\Gamma$ by 
$\mathcal{B}(\Gamma)$.

Another description of the measurable space $(\Gamma,\mathcal{B}(\Gamma))$ is 
also possible. 
For each $Y\in\mathcal{B}(X)$, let $\Gamma _Y$ be the space of 
all configurations contained in $Y$, $\Gamma _Y:=\left\{ \gamma \in \Gamma :
\left| \gamma \cap (X\!\setminus\! Y)\right| 
=0\right\}$,
and let $\Gamma _Y^{(n)}$ be the subset of all 
$n$-point configurations, $\Gamma _Y^{(n)}:=\left\{ \gamma \in \Gamma _Y:\left| \gamma \right|=n\right\}$, $n\in \N$, $\Gamma _Y^{(0)}:=\left\{ \emptyset \right\}$.
For $n\in\N$, there is a natural surjective mapping of 
\[
\widetilde{Y^n}:=\left\{ (x_1,...,x_n):x_i\in Y,x_i\neq x_j \hbox{ if }
i\neq j\right\}
\] 
onto $\Gamma _Y^{(n)}$ defined by
\begin{equation}
\begin{array}{ll}
\mathrm{sym}_Y^n:\widetilde{Y^n} & \rightarrow \Gamma _Y^{(n)} \\ 
(x_1,...,x_n) & \longmapsto \left\{ x_1,...,x_n\right\}
\end{array}
.  \label{1Eq1.2}
\end{equation}
This leads to a bijection between the space $\Gamma _Y^{(n)}$ and the 
symmetrization $\widetilde{Y^n}/S_n $ of $\widetilde{Y^n}$ under the 
permutation group $S_n$ over $\left\{ 1,...,n\right\} $, and then to a 
metrizable topology on $\Gamma _Y^{(n)}$. We denote the corresponding Borel 
$\sigma$-algebra on $\Gamma _Y^{(n)}$ by $\mathcal{B}(\Gamma _Y^{(n)})$. For 
$\Lambda\in\mathcal{B}(X)$ with compact closure 
($\Lambda\in\mathcal{B}_c(X)$ for short), one clearly has 
$\Gamma_\Lambda = \bigsqcup_{n=0}^\infty \Gamma_\Lambda^{(n)}$. In this case 
we endow $\Gamma_\Lambda$ with the topology of the disjoint union of 
topological spaces and with the corresponding Borel $\sigma$-algebra 
$\mathcal{B}(\Gamma_\Lambda)$ defined by the disjoint union of the 
$\sigma$-algebras $\mathcal{B}(\Gamma _\Lambda^{(n)})$, $n\in\N_0$, i.e.,
\[
\mathcal{B}(\Gamma_\Lambda)=\sigma\left( \left\{ \gamma\in\Gamma_\Lambda:
\vert \gamma\cap\Lambda'\vert = n\right\} \right),\quad 
\Lambda'\in\mathcal{B}_c(X), n\in\N_0. 
\]
The measurable space $(\Gamma ,\mathcal{B}(\Gamma ))$ is the projective limit 
of the measurable spaces $(\Gamma _\Lambda ,\mathcal{B}(\Gamma _\Lambda ))$, 
$\Lambda \in \mathcal{B}_c(X)$, with respect to the projections
\begin{equation}
\begin{array}{ll}
p_\Lambda :&\Gamma  \rightarrow \Gamma _\Lambda \\ 
&\gamma  \longmapsto \gamma_\Lambda:=\gamma\cap\Lambda
\end{array} . \label{1Eq1.4}
\end{equation}

Apart from the spaces described above we also consider the space of finite 
configurations
\[
\Gamma _0:=\bigsqcup_{n=0}^\infty \Gamma _X^{(n)}
\]
endowed with the topology of disjoint union of topological spaces and with 
the corresponding Borel $\sigma $-algebra denoted by $\mathcal{B}(\Gamma _0)$.

To define the $K$-transform, among the functions defined on $\Gamma_0$ we 
distinguish the space $B_{exp,ls}(\Gamma_0)$ of all complex-valued
exponentially bounded 
$\mathcal{B}(\Gamma_0)$-meas\-urable functions $G$ with local support, i.e., 
$G\!\!\upharpoonright _{\Gamma _0\backslash \Gamma_\Lambda}
\equiv 0$ for some $\Lambda \in \mathcal{B}_c(X)$
and there are $C_1,C_2 > 0$ such that
$|G(\eta)|\leq C_1 e^{C_2 |\eta|}$ for all $\eta \in \Gamma_0$. 
The $K$-transform of any 
$G\in B_{exp,ls}(\Gamma_0)$ is the mapping $KG:\Gamma\to\C$ defined at each 
$\gamma\in\Gamma$ by
\begin{equation}
(KG)(\gamma ):=\sum_{{\eta \subset \gamma}\atop{\vert\eta\vert < \infty} }
G(\eta ).
\label{Eq2.9}
\end{equation}
Note that for every $G\in B_{exp,ls}(\Gamma_0)$ the sum in (\ref{Eq2.9}) has 
only a finite number of summands different from zero, and thus $KG$ is a 
well-defined measurable cylinder function on $\Gamma$ with domain of
cylindricity $\Lambda$. Moreover, 
$|\left(KG\right)(\gamma)|\leq C_1 e^{(C_2+1)|\gamma_\Lambda|}$.

Throughout this work the so-called coherent states $e_\lambda(f)$ of 
$\mathcal{B}(X)$-meas\-urable functions $f$, defined by
$$
e_\lambda (f,\eta ):=\prod_{x\in \eta }f\left( x\right) ,\ \eta \in
\Gamma _0\!\setminus\!\{\emptyset\},\quad  e_\lambda (f,\emptyset ):=1,
$$
will play an essential role. This is partially due to the fact that 
the $K$-trans\-form of this class of functions coincides with the integrand 
functions of the Bogoliubov functionals (Section \ref{Section3}). More 
precisely, for every bounded $\mathcal{B}(X)$-measurable function $f$ with bounded support ($f\in B_{bs}(X)$ for short), one has $e_\lambda(f)\in B_{exp,ls}(\Gamma_0)$, and
\[
\left( Ke_\lambda (f)\right) (\gamma )=\prod_{x\in \gamma }(1+f(x)),\quad 
\gamma\in \Gamma.
\]

Besides the $K$-transform, we also consider the dual operator $K^*$. 
Let $\mathcal{M}_{\mathrm{fexp}}^1(\Gamma )$ denote the set of all probability 
measures $\mu$ on $(\Gamma ,\mathcal{B}(\Gamma ))$ with finite local 
exponential moments, i.e., 
\[
\int_\Gamma e^{\alpha|\gamma _\Lambda |}\,d\mu (\gamma )<\infty \quad 
\mathrm{for\,\,all}\,\,\Lambda \in \mathcal{B}_c(X)
\mathrm{\,\,and\,\,all\,\,} \alpha>0.
\]
By the definition of a dual operator, given a 
$\mu\in\mathcal{M}_{\mathrm{fexp}}^1(\Gamma )$, $K^*\mu =: \rho_\mu$ is a 
measure defined on $(\Gamma _0,\mathcal{B}(\Gamma _0))$ by
\begin{equation}
\int_{\Gamma _0}G(\eta )\,d\rho _\mu(\eta )=\int_\Gamma \left(
KG\right) (\gamma )\,d\mu (\gamma ),  \label{Eq2.16}
\end{equation}  
for all $G\in B_{exp,ls}(\Gamma_0)$. The measure $\rho_\mu$ is the correlation 
measure cor\-re\-spond\-ing to $\mu$. This definition shows, in particular, 
that $B_{exp,ls}(\Gamma_0)\subset L^1(\Gamma_0,\rho_\mu)$\footnote{Throughout 
this work all $L^p$-spaces, $p\geq 1$, consist of complex-valued functions.}. 
Moreover, on the dense set $B_{exp,ls}(\Gamma_0)$ in $L^1(\Gamma_0,\rho_\mu)$ 
the inequality 
$\Vert KG\Vert_{L^1(\mu)}\leq \Vert G\Vert_{L^1(\rho_\mu)}$ holds, allowing
an extension of the $K$-transform to a bounded operator 
$K:L^1(\Gamma_0,\rho_\mu)\to L^1(\Gamma,\mu)$ in such a way that equality 
(\ref{Eq2.16}) still holds for any $G\in L^1(\Gamma_0,\rho_\mu)$. For the 
extended operator the explicit form (\ref{Eq2.9}) still holds, now 
$\mu$-a.e. This means, in particular,
\begin{equation}
\left( Ke_\lambda (f)\right) (\gamma) = \prod_{x\in \gamma }(1+f(x)),\quad 
\mu \mathrm{-a.a.}\,\gamma\in\Gamma, 
\label{7.1}
\end{equation}  
for all $\mathcal{B}(X)$-measurable functions $f$ such that 
$e_\lambda(f)\in L^1(\Gamma_0,\rho_\mu)$.

\begin{remark}
\label{Rem1}All the notions described above as well as their relations are 
graphically summarized in the figure below. Having in mind the concrete 
application in Section \ref{Section5} below, let us mention the natural 
meaning of this figure in the context of an infinite particle system. The 
state of such a system is described by a probability measure $\mu$ on 
$\Gamma$ and the functions $F$ on $\Gamma$ are considered as observables of 
the system and they represent physical quantities which can be measured. The 
expected values of the measured observables 
correspond to the expectation values 
$\int_\Gamma F(\gamma)\,d\mu(\gamma)$. In this interpretation 
we call the functions 
$G$ on $\Gamma_0$ quasi-observables, because they are not observables
themselves, but can be used to construct observables via the $K$-transform.
In this way one obtains all observables which are additive in the particles,
e.g., number of particles, energy.
\end{remark}



\vspace{1truecm}

\begin{center}

\def\xpos {0}
\def\ypos {0}

\def\xlen {200}
\def\ylen {100}

\def\xinit {10}
\def\yinit {10}


\def\xlenline {180}           
\def\ylenline {80}           


\def\xposvectori {190}        
\def\yposvectori {90}        


\def\xletraposinit {-9}       
\def\xletraposend {191}       

\def\yletraposinit {-6}       
\def\yletraposend {91}       


\def\yexprtop {120}           
\def\yexprbottom {-18}        
\def\xexprleft {-14}          
\def\xexprright {203}         

\begin{picture} (\xlen, \ylen) (0, 0)


\put (\xinit,       \ypos) {\vector (1,0)  {\xlenline}}
\put (\xposvectori, \ypos) {\vector (-1,0) {\xlenline}}

\put (\xpos, \yinit)       {\vector (0,1) {\ylenline}}

\put (\xinit, \ylen)       {\vector (1,0) {\xlenline}}
\put (\xposvectori, \ylen) {\vector (-1,0) {\xlenline}}

\put (\xlen, \yposvectori) {\vector (0,-1) {\ylenline}}


\put (\xletraposinit, \yletraposinit) {\makebox(20,12)[t] {$G$}}
\put (\xletraposinit, \yletraposend)  {\makebox(20,12)[t]{$F$}}
\put (\xletraposend,  \yletraposend)  {\makebox(20,12)[t]{$\mu$}}
\put (\xletraposend,  \yletraposinit) {\makebox(20,12)[t]{$\rho _{_{}\mu}$}}


\put (\xpos, \yexprtop) 
{\makebox(\xlen, 12)[t] 
{$<F,\mu> = \displaystyle\int_{\Gamma} F(\gamma) d\mu (\gamma)$}}

\put (\xpos, \yexprbottom)
{\makebox(\xlen, 12)[t] {$<G,\rho _{_{}\mu}> = \displaystyle\int_{\Gamma_0} G(\eta) d\rho _{_{}\mu}(\eta)$}}

\put (\xexprleft, \ypos) 
{\makebox(12, \ylen)[l] {$K$}}

\put (\xexprright, \ypos) 
{\makebox(14, \ylen)[l] {$K^{*}$}}

\end{picture}

\end{center}

\vspace{1truecm}


On the underlying measurable space $(X,\mathcal{B}(X))$ let 
us consider a non-atomic Radon measure $\sigma$, i.e., $\sigma (\{x\})=0$ for 
any $x\in X$. The Poisson measure $\pi_\sigma$ with intensity $\sigma$ is the 
probability measure defined on $(\Gamma,\mathcal{B}(\Gamma))$ by 
\[
\int_\Gamma \exp \left( \sum_{x\in \gamma }\varphi (x)\right)\,
d\pi_\sigma(\gamma )
=\exp \left( \int_X\left( e^{\varphi (x)}-1\right)\,d\sigma(x)\right),
\quad \varphi\in \mathcal{D},
\]
where $\mathcal{D}:=C_0^\infty(X)$ denotes the Schwartz space of all 
infinitely differentiable real-valued functions on $X$ with compact support.
The correlation measure corresponding to the Poisson measure $\pi_\sigma$ is 
the so-called Lebesgue-Poisson measure 
$\lambda_\sigma$ (with intensity $\sigma$) 
$$
\lambda_\sigma:=\sum_{n=0}^\infty \frac{1}{n!} \sigma^{(n)},
$$
where each $\sigma^{(n)}$, $n\in \N$, is the symmetrization of the product 
measure $\sigma^{\otimes n}$, i.e., the image measure on 
$\Gamma_X^{(n)}$ of the measure $\sigma^{\otimes n}$ under the 
mapping $\mathrm{sym}_X^n$ defined in (\ref{1Eq1.2}). For $n=0$ we set 
$\sigma^{(0)}(\{\emptyset\}):=1$. The following Lebesgue-Poisson measure 
properties underline the importance of coherent states. 
First, 
$e_\lambda(f)\in L^p(\Gamma_0, \lambda_\sigma)$ whenever $f\in L^p(X,\sigma)$ 
for some $p\geq 1$, and, moreover,
 $\int_{\Gamma_0} \vert e_\lambda(f,\eta)\vert^p\,d\lambda_\sigma(\eta)
= \exp \left( \int_X \vert f(x)\vert^p\,d\sigma(x) \right)$.
Second, given a dense subspace $\mathcal{L}\subset L^2(X,\sigma)$, the set
$\{e_\lambda(f):f\in \mathcal{L}\}$ is total in $L^2(\Gamma_0,\lambda_\sigma)$.

\section{Bogoliubov functionals\label{Section3}}

For the case $X=\R^d$, $d\in\N$, we refer to \cite{Naz85} and his own 
references therein.

\begin{definition}
\label{9def1} Let $\mu$ be a probability measure on 
$(\Gamma ,\mathcal{B}(\Gamma ))$. The Bogoliubov functional $L_\mu$ 
corresponding to $\mu$ is a functional defined at each 
$\mathcal{B}(X)$-measurable function $\theta$ by 
\[
L_\mu (\theta ):=\int_\Gamma \prod_{x\in \gamma }(1+\theta (x))d\mu
(\gamma ), 
\]
provided the right-hand side exists for $|\theta|$.
\end{definition}

We note that if $L_\mu(|\theta|)< \infty$, then the product  
$\prod_{x\in\gamma} (1+\theta(x))$ is $\mu$-a.e.~absolutely convergent. For 
the definition and properties of infinite products see \cite{Kn64}.

It is clear that the domain of a Bogoliubov functional $L_\mu$ depends on the 
measure $\mu$ fixed on $(\Gamma, \mathcal{B} (\Gamma))$. Conversely, the 
domain of a Bogoliubov functional reflects special properties over the 
underlying measure on $(\Gamma, \mathcal{B} (\Gamma))$. For instance, 
probability measures $\mu$ for which the Bogoliubov functional is well-defined 
on multiples of indicator functions $1\!\!1_\Lambda$, 
$\Lambda\in \mathcal{B}_c(X)$, necessarily have finite local exponential 
moments, i.e., $\mu\in\mathcal{M}_{\mathrm{fexp}}^1(\Gamma)$. In fact, for 
all $\alpha>0$ and all $\Lambda \in \mathcal{B}_c(X)$ we find 
$$
\int_{\Gamma }e^{\alpha\left| \gamma_\Lambda \right|}\,d\mu (\gamma)
=\int_{\Gamma}\prod_{x\in \gamma}e^{\alpha 1\!\!1_\Lambda (x)}\,d\mu (\gamma )
=L_\mu((e^\alpha-1)1\!\!1_\Lambda ) < \infty.
$$

In the sequel, for each probability measure $\mu$ on 
$(\Gamma ,\mathcal{B}(\Gamma ))$ and each $\Lambda\in\mathcal{B}_c(X)$, we 
denote by $\mu^\Lambda:= \mu\circ (p_\Lambda )^{-1}$ the image measure on 
$\Gamma_\Lambda$ of the measure $\mu$ under the projection $p_\Lambda$ 
defined in (\ref{1Eq1.4}), i.e., $\mu^\Lambda$ is the projection of $\mu$ onto 
$\Gamma_\Lambda$. Given a $\Lambda\in\mathcal{B}_c(X)$, the definition of a
Bogoliubov functional $L_\mu$ on the space of all functions $\theta$ with 
support contained in $\Lambda$ reduces to the Bogoliubov functional 
$L_{\mu^\Lambda}$:  
\[
L_\mu(\theta ) =\int_\Gamma \prod_{x\in \gamma }(1+\theta (x))d\mu
(\gamma ) 
=\int_\Gamma \prod_{x\in \gamma _\Lambda }(1+\theta (x))d\mu (\gamma )
=L_{\mu^\Lambda }(\theta ).
\]
Furthermore, one may straightforwardly express the $\mu$-measure of a 
large class of sets by the Bogoliubov functional $L_\mu$. In fact, given 
$z_1,...,z_n\in\C$ and a collection of mutually disjoint sets 
$\Lambda_1,...,\Lambda _n\in \mathcal{B}_c(X)$, 
$\Delta :=\bigsqcup _{i=1}^n\Lambda _i$, $n\in\N$, the above computation has 
shown that 
\[
L_\mu\left( \sum_{i=1}^nz_i1\!\!1_{\Lambda _i}-1\!\!1_\Delta \right)
=\int_{\Gamma }\prod_{x\in \gamma_\Delta }\left(
\sum_{i=1}^nz_i1\!\!1_{\Lambda _i}(x)\right) d\mu(\gamma ).
\]
Since $\Gamma_\Delta$ may be written as the disjoint union 
\[
\Gamma _\Delta =\bigsqcup_{k_1,...,k_n=0}^\infty \left\{ \gamma \in \Gamma
_\Delta :\left| \gamma _{\Lambda _i}\right| =k_i,i=1,...,n\right\} , 
\]
the latter integral is then equal to 
\[
\sum_{k_1,...,k_n=0}^\infty z_1^{k_1}...z_n^{k_n}\mu \left(
\left\{ \gamma \in \Gamma :\left| \gamma _{\Lambda _i}\right|
=k_i,i=1,...,n\right\} \right) .
\]
Heuristically, this means that 
\begin{eqnarray}
&&\mu \left( \left\{ \gamma \in \Gamma :\left| \gamma
_{\Lambda _i}\right| =k_i,i=1,...,n\right\} \right)  \label{9Eq1.1} \\
&=&\frac 1{k_1!...k_n!}\frac{\partial ^{k_1+...+k_n}}{\partial
z_1^{k_1}...\partial z_n^{k_n}}L_{_{}\mu }\left(
\sum_{i=1}^nz_i1\!\!1_{\Lambda _i}-1\!\!1_{\bigcup_{i=1}^n\Lambda _i}\right) 
\Biggl\vert_{z_1=...=z_n=0}.  \nonumber
\end{eqnarray}

According to the definition of the $\sigma$-algebra $\mathcal{B}(\Gamma)$, the 
collection of sets appearing in the left-hand side of the informal equality 
(\ref{9Eq1.1}) already characterizes the measure $\mu$. 

Of course, in order to apply the above procedure we must assume 
that the Bogoliubov functional $L_\mu$ is well-defined and 
differentiable on the class of linear combinations of indicator functions 
which appears in (\ref{9Eq1.1}). As the linear space spanned by indicator 
functions or the spaces of measurable functions are both difficult to handle, 
throughout this work we will consider Bogoliubov functionals on a 
$L^1(X,\sigma)=:L^1(\sigma)$ space, for some Radon measure $\sigma$ defined on 
the space $(X,\mathcal{B}(X))$. Furthermore, we will assume that the 
Bogoliubov functionals are entire. We observe that from the viewpoint of 
particle systems these restrictions are natural. Actually, even stronger 
properties should be expected.

In the sequel, we fix on $(X,\mathcal{B}(X))$ a non-atomic Radon measure 
$\sigma$ which we assume to be non-degenerate, i.e., $\sigma(O)>0$ for all 
non-empty open sets $O\subset X$, and, in addition, $\sigma(X)=\infty$. 

We recall that a functional $A:L^1(\sigma)\to\C$ is entire on $L^1(\sigma)$ 
whenever $A$ is locally bounded, and for all $\theta_0,\theta\in L^1(\sigma)$ 
the mapping $\C\ni z\mapsto A(\theta_0 + z\theta)\in\C$ is entire. Thus, at 
each $\theta_0\in L^1(\sigma)$, every entire functional $A$ on $L^1(\sigma)$ 
has a representation in terms of its Taylor expansion,
$$
A(\theta_0+ z \theta)=\sum_{n=0}^\infty \frac{z^n}{n!}
d^nA(\theta_0;\theta ,...,\theta),\quad z\in\C, \theta\in L^1(\sigma),
$$           
see e.g.~\cite{Ba85}, \cite{Di81}. The next theorem states 
a characterization result for the differentials $d^nA(\theta_0;\cdot)$ of an 
entire functional $A$ on $L^1(\sigma)$.

\begin{theorem}
\label{9Prop9.1.1}Let $A$ be an entire functional on $L^1(\sigma )$. Then
each differential $d^nA(\theta _0;\cdot), n\in\N, \theta _0\in L^1(\sigma)$ is 
defined by a (symmetric) kernel in $L^\infty (X^n,\sigma ^{\otimes n})$ 
denoted by 
$\frac{\delta ^nA(\theta _0)}{\delta \theta _0(x_1)...\delta \theta _0(x_n)}$ 
and called the variational derivative of $n$-th order of $A$ at the 
point $\theta _0$. In other words, 
\begin{eqnarray*}
d^nA(\theta _0;\theta _1,...,\theta _n) &:=&\frac{\partial ^n}{\partial
z_1...\partial z_n}A\left( \theta _0+\sum_{i=1}^nz_i\theta _i\right) %
\Big\vert_{z_1=...=z_n=0} \\
&\hbox{\rm{=:}}&\int_{X^n}\frac{\delta ^nA(\theta _0)}{\delta \theta
_0(x_1)...\delta \theta _0(x_n)}\prod_{i=1}^n\theta _i(x_i)d\sigma
^{\otimes n}(x_1,..,x_n)
\end{eqnarray*}
for all $\theta _1,...,\theta _n\in L^1(\sigma )$. Moreover, for all $r>0$
\begin{equation}
\left\| \frac{\delta^n A(\theta _0)}{\delta \theta
_0(x_1)...\delta \theta _0(x_n)}\right\|_{L^\infty(X^n,\sigma^n)}
\leq n! \left(\frac{e}{r}\right)^n \sup_{\|\theta^\prime \|_{L^1(\sigma)} \leq r} |A(\theta_0+\theta^\prime)| \label{Duarte}
\end{equation}
\end{theorem}

\begin{remark}
\label{Verdasca}
According to Theorem \ref{9Prop9.1.1}, the Taylor expansion of an entire 
functional $A$ at a point $\theta _0\in L^1(\sigma)$ may be written in the 
form 
\[
A(\theta _0+\theta ) =\sum_{n=0}^\infty \frac 1{n!}\int_{X^n}
\frac{\delta ^nA(\theta _0)}{\delta \theta _0(x_1)...\delta\theta _0(x_n)}\prod_{i=1}^n\theta (x_i)d\sigma ^{\otimes n}(x_1,...,x_n),
\]
for all $\theta\in L^1(\sigma)$. Using the notation  
\[
\left(D^{\left|\eta\right|} A\right)(\theta_0;\eta)
:=\frac{\delta ^nA(\theta _0)}{\delta \theta _0(x_1)...\delta \theta
_0(x_n)}\quad \mathit{for}\mathrm{\,\,}\eta =\{x_1,...,x_n\}\in \Gamma
_X^{(n)},n\in\N ,
\]
this means
$$
A(\theta _0+\theta )=\int_{\Gamma _0}e_\lambda (\theta ,\eta )
\left(D^{\left|\eta\right|} A\right)(\theta_0;\eta)
d\lambda_\sigma(\eta ).
$$
Concerning the estimate (\ref{Duarte}), we note that the entire property of 
$A$ does not insure that for every $r>0$ the supremum on the right-hand side 
is always finite. This will hold if, in addition, the entire functional $A$ is 
of bounded type, that is,
\[
\forall\,r>0,\ 
\sup_{\Vert\theta\Vert_{L^1(\sigma)}\leq r}\left|A(\theta_0+\theta)\right|<\infty,\ \forall\,\theta_0\in L^1(\sigma).
\] 
For simplicity, throughout this work we will assume this assumption.
\end{remark}

The proof of the first part of this result is of a technical nature outside of 
the present context. However, it contains a few steps which we will need to 
prove the second part. Because of this, we just present a sketch of the proof 
conveniently adapted to our aims and complemented with suitable references for 
a detailed proof.  

\medskip

\noindent
\textbf{Proof.} According to the Cauchy formula for analytic functions
on Banach spaces, each differential $d^nA(\theta _0;\cdot)$ 
of an entire functional $A$ on $L^1(\sigma)$ is a bounded symmetric $n$-linear
functional on $L^1(\sigma )$. In particular, for $n=1$, the first order 
differential $dA(\theta _0;\cdot)$ is a bounded linear functional on 
$L^1(\sigma)$, insuring that it can be represented by a kernel in 
$L^\infty (\sigma)$, the so-called first variational derivative 
$\frac{\delta A(\theta _0)}{\delta \theta _0(x)}$. Furthermore, the (usual) 
operator norm of the bounded linear functional $dA(\theta _0;\cdot)$ is equal 
to $\left\| \frac{\delta A(\theta _0)}{\delta \theta _0(\cdot )}\right\| _{L^\infty(X,\sigma )}$.

For higher orders, the proof of existence of the corresponding variational
derivatives is a straightforward consequence of the isometries between the 
Banach spaces 
\begin{equation}
B_n(L^1(X,\sigma ))\simeq \left( L^1(X^n,\sigma ^{\otimes n})\right)
^{\prime }\simeq L^\infty (X^n,\sigma ^{\otimes n}),  \label{9Eq1.7}
\end{equation}
$B_n(L^1(X,\sigma ))$ being the space of all bounded $n$-linear
functionals on $L^1(X,\sigma )$. For the proof see e.g.~\cite{DiU79}, 
\cite{Sch71}, \cite{Trv67}. These isometries prove, on the one hand, the 
existence of the variational derivatives 
$\frac{\delta ^nA(\theta _0)}{\delta \theta _0(x_1)...\delta 
\theta _0(x_n)}\in L^\infty (X^n,\sigma ^{\otimes n})$ as kernels for 
$d^nA(\theta_0;\cdot)$, and, on the other hand, that the operator norm of 
$d^nA(\theta _0;\cdot)\in B_n(L^1(X,\sigma ))$ is given by
$\left\| \frac{\delta ^nA(\theta _0)}{\delta \theta _0(\cdot ) ... 
\delta\theta_0(\cdot)}\right\| _{L^\infty(X^n,\sigma ^{\otimes n})}$.
This shows the first part of the theorem. To prove the second one, we observe 
that by the Cauchy formula, for any $\theta \in L^1(\sigma )$ one has 
\[
\frac 1{n!}d^nA(\theta_0;\theta,...,\theta )
=\frac 1{2\pi i}\int_{\left| z\right| =r}
\frac{A(\theta_0+z\theta )}{z^{n+1}}dz
\]
for any $r>0$ and any $n\in\N$. Therefore 
\[
\left| d^nA(\theta_0;\theta,...,\theta)\right| 
\leq n!\sup_{\left\| \theta ^{\prime
}\right\| _{L^1(\sigma )}\leq r} \left| A(\theta_0+ \theta ^{\prime
})\right| \left( \frac{\left\| \theta \right\| _{L^1(\sigma )}}r\right) ^n, 
\]
and an application of the polarization identity extends this 
inequality to $\theta _1,...,\theta _n\in L^1(\sigma )$: 
\[
\left| d^nA(\theta_0;\theta _1,...,\theta _n)\right|
\leq n!\left( \frac er\right) ^n\sup_{\left\| \theta ^{\prime }\right\|
_{L^1(\sigma )}\leq r} \left| A(\theta_0+\theta ^{\prime })\right|
\prod_{i=1}^n\left\| \theta _i\right\| _{L^1(\sigma )},
\]
see e.g.~\cite[Theorem 1.7]{Di81}.\hfill$\blacksquare$

\begin{remark}
Observe that the first isometry in (\ref{9Eq1.7}) is specific of $L^1$ spaces. 
The analogous result does not hold neither for other $L^p$-spaces, nor Banach 
spaces of continuous functions, or Sobolev spaces.
\end{remark}

Theorem \ref{9Prop9.1.1} stated for Bogoliubov functionals yields the next
result. In particular, it gives a rigorous sense to the discussion at the 
beginning of this section.

\begin{corollary}
\label{9Prop9.1.2}Let $L_\mu$ be a Bogoliubov functional corresponding to 
some probability measure $\mu$ on $(\Gamma ,\mathcal{B}(\Gamma))$. If $L_\mu$ 
is entire of bounded type on $L^1(\sigma)$, then the measure $\mu$ is locally 
absolutely continuous with respect to the Poisson measure $\pi_\sigma$, i.e., 
for all $\Lambda\in \mathcal{B}_c(X)$ the measure 
$\mu^\Lambda= \mu\circ (p_\Lambda )^{-1}$ is absolutely continuous with 
respect to $\pi_\sigma^\Lambda=\pi_\sigma\circ (p_\Lambda )^{-1}$. Moreover, 
for all $\Lambda \in \mathcal{B}_c(X)$ one has 
\[
\frac{d\mu ^\Lambda }{d\pi_\sigma^\Lambda }(\gamma )=e^{\sigma
(\Lambda )}
\left(D^{\left|\gamma\right|} L_\mu\right)(-1\!\!1_\Lambda;\gamma)\quad 
\mathit{for}
\mathrm{\,\,}\pi _\sigma^\Lambda \mathrm{-\mathit{a.a.}\,\,}\gamma \in
\Gamma _\Lambda ,
\]
and for each $r>0$ there exists a constant $C\geq 0$ such that 
\[
\left|\frac{d\mu ^\Lambda }{d\pi_\sigma^\Lambda }(\gamma )\right|
\leq e^{\sigma(\Lambda)} C |\gamma|! \left(\frac{e}{r}\right)^{|\gamma|}
\quad \text{for }\pi_\sigma^\Lambda\text{-a.a. }
\gamma \in \Gamma^{(n)}_\Lambda .
\]
\end{corollary}

\noindent
\textbf{Proof.} In Theorem \ref{9Prop9.1.1} replace $A$ by the functional 
$L_\mu$ and $\theta _0$ by an indicator function $-1\!\!1_\Lambda $ for some 
$\Lambda \in \mathcal{B}_c(X)$. Thus, for all functions 
$\theta \in L^1(\sigma )$ with support contained in $\Lambda$, we find
\begin{eqnarray*}
L_\mu(\theta ) &=&L_{_{}\mu }(-1\!\!1_\Lambda +(\theta +1\!\!1_\Lambda)) \\
&=&\int_{\Gamma_\Lambda} \prod_{x\in \eta }(1+\theta (x))
\left(D^{\left|\eta\right|} L_\mu\right)(-1\!\!1_\Lambda;\eta)
d\lambda _\sigma(\eta ).
\end{eqnarray*}
On the other hand, according to the considerations done at the beginning of
this section, we also have 
\[
L_\mu(\theta )=\int_{\Gamma _\Lambda }\prod_{x\in \gamma }(1+\theta
(x))d\mu ^\Lambda (\gamma ). 
\]
Therefore 
\[
\int_{\Gamma _\Lambda }\prod_{x\in \gamma }(1+\theta (x))d\mu ^\Lambda
(\gamma )=\int_{\Gamma _\Lambda }\prod_{x\in \eta }(1+\theta (x))
\left(D^{\left|\eta\right|} L_\mu\right)(-1\!\!1_\Lambda;\eta)
d\lambda_\sigma(\eta )
\]
for all functions $\theta \in L^1(\sigma )$ with support contained in 
$\Lambda$. The proof follows by a monotone class argument.\hfill$\blacksquare \medskip$

Since $\mu\in\mathcal{M}_{\mathrm{fexp}}^1(\Gamma )$ whenever the 
corresponding Bogoliubov functional is well-defined on the whole space
$L^1(\sigma)$, one can associate the correlation measure $\rho_\mu=K^{*}\mu$ 
to a such measure. Equalities (\ref{7.1}) and (\ref{Eq2.16}) then yield a 
description of the functional $L_\mu$ in terms of the measure $\rho_\mu$:
\begin{equation}
L_\mu(\theta) 
=\int_\Gamma \left( Ke_\lambda(\theta )\right) (\gamma )d\mu (\gamma)
=\int_{\Gamma _0}e_\lambda(\theta ,\eta )d\rho_\mu(\eta ).
\label{9Eq1.12}
\end{equation}
Within this formalism Theorem \ref{9Prop9.1.1} states as follows.

\begin{proposition}
\label{9Prop9.1.3}Let $L_\mu$ be an entire Bogoliubov functional of bounded 
type on $L^1(\sigma )$. Then the measure $\rho_\mu$ is absolutely continuous
with respect to the Lebesgue-Poisson measure $\lambda_\sigma$ and
the Radon-Nykodim derivative $k_\mu:=\frac{d\rho_\mu}{d\lambda_\sigma}$ is 
given by 
\[
k_\mu(\eta )=
\left(D^{\left|\eta\right|} L_\mu\right)(0;\eta) 
\quad \mathit{for}\mathrm{\,\,}
\lambda_\sigma \mathrm{-\mathit{a.a.}\,\,}\eta \in \Gamma _0. 
\]
Furthermore, for each $r>0$ there is a constant $C\geq 0$ such that 
\[
\left| 
\left(D^{\left|\eta\right|} L_\mu\right)(0;\eta)
\right| \leq C\left| \eta \right| !\left( \frac
er\right) ^{\left| \eta \right| }\quad \mathit{for}\mathrm{\,\,}\lambda_\sigma \mathrm{-\mathit{a.a.}}\text{ }\eta \in \Gamma _0. 
\]
\end{proposition}

In the sequel we call $k_\mu$ the correlation function corresponding to $\mu$. 

\medskip

\noindent
\textbf{Proof.} A straightforward application of Theorem \ref{9Prop9.1.1} 
yields
\[
L_\mu(\theta )=\int_{\Gamma _0}e_\lambda(\theta ,\eta )
\left(D^{\left|\eta\right|} L_\mu\right)(0;\eta)
d\lambda_\sigma (\eta ),\quad \theta \in L^1(\sigma ),
\]
and
\[
\left| 
\left(D^{\left|\eta\right|} L_\mu\right)(0;\eta)
\right| \leq C\left| \eta \right| !\left( \frac
er\right) ^{\left| \eta \right| },\quad \lambda_\sigma \mathrm{-a.a.}
\text{ }\eta \in \Gamma _0, 
\]
for some $C\geq 0$ depending on $r$. Expression (\ref{9Eq1.12}) 
then allows to identify $k_\mu(\eta)$ with $\left(D^{\left|\eta\right|} L_\mu\right)(0;\eta)$.\hfill$\blacksquare$

\begin{remark}
\label{Rem12}
Proposition \ref{9Prop9.1.3} shows that the correlation functions 
$k_\mu^{(n)}:=k_\mu\!\!\upharpoonright _{\Gamma_X^{(n)}}$ are the Taylor 
coefficients of the Bogoliubov functional $L_\mu$. In other words, $L_\mu$ is 
the generating functional for the correlation functions $k_\mu^{(n)}$. This 
was also the reason why N.~N.~Bogoliubov introduced these functionals. 
Furthermore, Bogoliubov functionals are also related to the general infinite 
dimensional analysis on configuration spaces, cf.~e.g.~\cite{KoKuOl00}. 
Namely, through the unitary isomorphism $S_\lambda$ defined in \cite{KoKuOl00} 
between the space $L^2(\Gamma_0,\lambda_\sigma)$ and the Bargmann-Segal space 
one has $L_\mu=S_\lambda(k_\mu)$. 
\end{remark}

\begin{proposition}
\label{9Prop9.1.5}For any Bogoliubov functional $L_\mu$ entire of bounded type 
on $L^1(\sigma)$ the following relations between variational derivatives hold: 
\begin{equation}
\left(D^{\left|\eta\right|} L_\mu\right)(\theta;\eta)\!
=\!\int_{\Gamma _0}k_\mu (\eta \cup \xi )e_\lambda (\theta,\xi)
d\lambda_\sigma(\xi )\quad \mathit{for}\mathrm{\,\,}\lambda_\sigma
\mathrm{-\mathit{a.a.}\,\,}\eta \in \Gamma _0  \label{9Eq1.2}
\end{equation}
and, more generally, 
\[
\left(D^{\left|\eta\right|} L_\mu\right)(\theta _1+\theta _2;\eta)\!
=\!\int_{\Gamma _0}\!\!
\left(D^{\left|\eta\cup \xi\right|} L_\mu\right)\!(\theta _1;\eta\cup \xi)
e_\lambda (\theta _2,\xi )d\lambda_\sigma (\xi)\quad 
\mathit{for}\mathrm{\,\,}\lambda_\sigma\mathrm{-\mathit{a.a.}\,\,}\eta \in 
\Gamma _0, 
\]
for $\theta ,\theta _1,\theta _2\in L^1(\sigma)$.
\end{proposition}

To prove this result as well as other forthcoming ones the next lemma shows 
to be useful.

\begin{lemma}
\label{Lmm1}(\cite{FicFre91}, \cite{KoKuOl00}, \cite{Ru69}) The following
equality holds 
\[
\int_{\Gamma _0}\int_{\Gamma _0}G(\eta \cup \xi )H(\xi ,\eta )d\lambda
_\sigma (\eta )d\lambda_\sigma (\xi )=\int_{\Gamma _0}G(\eta)
\sum_{\xi \subset \eta }H(\xi ,\eta \backslash \xi )d\lambda_\sigma(\eta ) 
\]
for all positive measurable functions $G:\Gamma _0\to\R$
and $H:\Gamma _0\times \Gamma _0\to\R$.
\end{lemma}

\noindent
\textbf{Proof.} According to Theorem \ref{9Prop9.1.1}, for all 
$\theta_1,\theta_2,\theta \in L^1(\sigma)$ one has
\[
L_\mu(\theta_1+\theta_2+\theta) = \int_{\Gamma_0} 
\left(D^{|\eta|}L_\mu\right)(\theta_1+\theta_2;\eta)
\ e_\lambda(\theta,\eta) d\lambda_\sigma(\eta)
\]
as well as
\[
L_\mu(\theta_1+\theta_2+\theta) = \int_{\Gamma_0} \left(D^{|\eta|}L_\mu\right)(\theta_1;\eta)
\ e_\lambda(\theta_2+\theta,\eta) d\lambda_\sigma(\eta).
\]
The bounds obtained in Theorem \ref{9Prop9.1.1} allows to apply Lemma 
\ref{Lmm1} to the latter equality yielding
\[
\int_{\Gamma_0}\int_{\Gamma_0} \left(D^{|\eta\cup\xi|}L_\mu\right)(\theta_1;\eta\cup \xi)
\ e_\lambda(\theta_2,\xi) d\lambda_\sigma(\xi) e_\lambda(\theta,\eta) d\lambda_\sigma(\eta).
\]
The second stated equality follows by a monotone class argument. By 
Proposition \ref{9Prop9.1.3} one sees that (\ref{9Eq1.2}) is a special case of 
the derived result for $\theta_1=0$ and $\theta_2=\theta$.\hfill $\blacksquare \medskip$

A particular application of Proposition \ref{9Prop9.1.5} yields the next two 
formulas well-known in statistical mechanics, see e.g.~\cite{Ru70}, and in 
the theory of point processes, see e.g.~\cite{DaVJ88}.

\begin{corollary}
\label{9Prop9.1.6}Under the conditions of Proposition \ref{9Prop9.1.5}, for 
all $\Lambda \in \mathcal{B}_c(X)$ we have 
\begin{equation}
k_\mu(\eta )=\int_{\Gamma _\Lambda }\frac{d\mu ^\Lambda }{d\pi_\sigma^\Lambda }
(\eta \cup \gamma )d\pi _\sigma^\Lambda (\gamma
)\quad \mathit{for}\mathrm{\,\,}\lambda _\sigma\mathrm{-\mathit{a.a.}%
\,\,}\eta \in \Gamma _\Lambda, \label{M.5} 
\end{equation}
and 
\begin{equation}
\frac{d\mu ^\Lambda }{d\pi _\sigma ^\Lambda }(\gamma )=e^{\sigma
(\Lambda )}\int_{\Gamma _\Lambda }(-1)^{|\eta |}k_\mu(\gamma \cup \eta
)d\lambda _\sigma (\eta )\quad \mathit{for}\mathrm{\,\,}\pi _\sigma
^\Lambda \mathrm{-\mathit{a.a.}\,\,}\gamma \in \Gamma _\Lambda . \label{M.6}
\end{equation}
\end{corollary}

\noindent
\textbf{Proof.} Fixing a $\Lambda \in \mathcal{B}_c(X)$, in Proposition 
\ref{9Prop9.1.5} replace both functions $\theta $ and $\theta _1$ by the 
function $-1\!\!1_\Lambda $ and $\theta _2$ by $1\!\!1_\Lambda $. The 
expressions for the densities given in Corollary \ref{9Prop9.1.2} and 
Proposition \ref{9Prop9.1.3} complete the proof.\hfill $\blacksquare \medskip$

\begin{remark}
\label{M.4}Corollary \ref{9Prop9.1.6} may be stated under more general 
conditions. Giv\-en a probability measure $\mu$ on 
$(\Gamma ,\mathcal{B}(\Gamma))$ such that $\int_\Gamma |\gamma _\Lambda |^n\,d\mu (\gamma ), \int_{\Gamma _\Lambda }2^{|\eta |}d\rho_\mu(\eta)<\infty$ for all 
$\Lambda\in\mathcal{B}_c(X)$ and all $n\in \N$, one can show that $\mu$ is locally absolutely 
continuous with respect to the Poisson measure $\pi_\sigma$ if and only if the 
correlation measure $\rho_\mu$ 
is absolutely continuous with respect to the Lebesgue-Poisson measure 
$\lambda_\sigma$. Under these conditions, equalities (\ref{M.5}) and 
(\ref{M.6}) hold (see e.g.~\cite{KoKu99}). 
\end{remark}

\begin{corollary}
\label{9Prop9.1.4}Let $L_\mu$ be an entire Bogoliubov functional of bounded 
type on $L^1(\sigma )$. For any $\mathcal{B}(\Gamma_0)$-measurable 
function $G:\Gamma_0\to\R$ such that there is a $f\in L^1(\sigma )$ with 
$\left| G\right| \leq e_\lambda(f)$, one has
\[
\int_{\Gamma_0}G(\eta )
\left(D^{|\eta|} L_\mu\right)(\theta;\eta)\,
d\lambda_\sigma(\eta ) 
= \int_{\Gamma _0}\sum_{\xi \subset \eta}G(\xi
)e_\lambda(\theta ,\eta \backslash \xi )\,d\rho_\mu(\eta ),
\]
for all $\theta \in L^1(\sigma )$.
\end{corollary}

According to Proposition \ref{9Prop9.1.3}, the correlation function $k_\mu$ 
of an entire Bogoliubov functional on $L^1(\sigma)$ fulfills the so-called 
generalized Ruelle bound, that is, for any $0\leq\varepsilon\leq 1$ and any 
$r>0$ there is some constant $C\geq 0$ depending on $r$ such that 
\begin{equation}
k_\mu(\eta )\leq C\left(\left|\eta\right|!\right)^{1-\varepsilon} 
\left( \frac er\right) ^{\left|\eta\right|},\quad 
\lambda _\sigma \mathrm{-a.a.}\text{ }\eta \in\Gamma _0.\label{KuRu}
\end{equation}
In our case, $\varepsilon$ is zero. We note that if (\ref{KuRu}) holds for 
$\varepsilon =1$ and for at least one $r>0$, then condition (\ref{KuRu}) is 
the classical Ruelle bound. For a general $0<\varepsilon\leq 1$ one may state 
the following result. 

\begin{proposition}
\label{9Prop9.1.10}(\cite{KoKu99c}) If there are a function 
$0\leq C\in L_{\mathrm{loc}}^1(X,\sigma )$ and a $0<\varepsilon\leq 1$ such 
that 
$$
k_\mu(\eta )\leq \left( \left| \eta \right| !\right) ^{1-\varepsilon}
e_\lambda(C,\eta ),\quad \lambda _\sigma \mathrm{-\mathit{a.a.}\,\,}
\eta \in \Gamma _0,  
$$
then there are constants $c_1=c_1(\varepsilon), c_2=c_2(\varepsilon)>0$ such 
that
$$
\left| L_\mu(\varphi )\right| \leq c_1\exp \left( \left\| \varphi
\right\| _{L^1(c_2C\sigma )}^{1/\varepsilon }\right) ,\quad 
\varphi\in\mathcal{D}.
$$
Furthermore, $L_\mu$  is an entire functional of bounded type on 
$L^1(C\sigma )$.
\end{proposition}

The definition of a Bogoliubov functional clearly shows that for any 
probability measure
$\mu \in \mathcal{M}_{\mathrm{fexp}}^1(\Gamma )$ $L_\mu$ is a normalized 
functional, that is, $L_\mu(0)=1$. If, in addition, $L_\mu$ is an entire 
functional on $L^1(\sigma )$, then, according to Corollary \ref{9Prop9.1.2},
for all $\Lambda\in\mathcal{B}_c(X)$ we have  
\[
\left(D^{\left|\gamma\right|} L_\mu\right)(-1\!\!1_\Lambda;\gamma)
=e^{-\sigma (\Lambda )}\frac{d\mu
^\Lambda }{d\pi _\sigma^\Lambda }(\gamma )\geq 0,\quad 
\lambda_\sigma-\mathrm{a.a.\,\,}\gamma \in \Gamma _\Lambda.
\]
These conditions are also sufficient to insure that a generic entire
functional on $L^1(\sigma )$ is a Bogoliubov functional corresponding to
some measure in $\mathcal{M}_{\mathrm{fexp}}^1(\Gamma )$.

\begin{proposition}
\label{9Prop9.1.11}Let $L$ be a normalized entire functional of bounded type 
on $L^1(\sigma)$ such that for all $\Lambda \in \mathcal{B}_c(X)$
\[
\left(D^{\left|\eta\right|} L\right)(-1\!\!1_\Lambda;\eta)\geq 0,\quad 
\lambda _\sigma \mathrm{-\mathit{a.a.}\,\,}\eta \in \Gamma_\Lambda.
\]
Then there is a unique probability measure 
$\mu \in \mathcal{M}_{\mathrm{fexp}}^1(\Gamma )$ such that for all 
$\theta \in L^1(\sigma )$
\begin{equation}
L(\theta )
=\int_\Gamma \prod_{x\in \gamma }(1+\theta (x))d\mu (\gamma).\label{Ku22}
\end{equation}
\end{proposition}

\noindent
\textbf{Proof.} For any $\Lambda \in \mathcal{B}_c(X)$ let us define the 
function 
\[
G_\Lambda (\eta ):=\left(D^{\left|\eta\right|} L\right)(-1\!\!1_\Lambda;\eta)
\geq 0,\quad \eta \in \Gamma_\Lambda.
\]
For all $\Lambda \in \mathcal{B}_c(X)$ we have
\begin{eqnarray*}
\int_{\Gamma _\Lambda }G_\Lambda (\eta )d\lambda _\sigma (\eta )
&=&\int_{\Gamma _0}e_\lambda(1\!\!1_\Lambda ,\eta )
\left(D^{\left|\eta\right|} L\right)(-1\!\!1_\Lambda;\eta)
d\lambda _\sigma (\eta ) \\
&=&L(1\!\!1_\Lambda -1\!\!1_\Lambda )=L(0)=1,
\end{eqnarray*}
allowing to define a family of probability measures $\mu^\Lambda$ on 
$(\Gamma _\Lambda ,\mathcal{B}(\Gamma_\Lambda ))$ by
\[
\mu ^\Lambda (A):=\int_{\Gamma _\Lambda }1\!\!1_A(\eta )G_\Lambda (\eta
)d\lambda _\sigma (\eta ),\quad A\in \mathcal{B}(\Gamma _\Lambda ). 
\]
Similarly, one verifies that the family 
$(\mu^\Lambda)_{\Lambda \in \mathcal{B}_c(X)}$ is consistent. Therefore, by 
the version of the Kolmogorov theorem for the projective limit space 
$(\Gamma,\mathcal{B}(\Gamma))$ \cite[Chapter V, Theorem~5.1]{P67}, there is a 
unique probability measure $\mu$ on $\Gamma$ such that the measures 
$\mu^\Lambda$ are the projections of $\mu$. From the definition of $G_\Lambda$
follows the relation (\ref{Ku22}) between $L$ and $\mu$ for every $\theta$ 
supported in $\Lambda$. The $L^1$-continuity of $L$ and monotone convergence 
arguments extend this relation to all non-negative functions 
$\theta \in L^1(\sigma)$. The general relation follows from dominated 
convergence results.\hfill $\blacksquare$

\section{Bogoliubov equations\label{Section4}}

Particularly interesting is the characterization of Gibbs measures through 
the Bogoliubov functionals. 

Given a pair potential $\phi:X\times X\to\R\cup\{+\infty\}$, that is, a 
symmetric measurable function, let $E:\Gamma_0\to\R\cup\{+\infty\}$ be the 
energy functional and $W:\Gamma_0\times \Gamma \to\R\cup\{+\infty\}$ be the 
interaction energy defined for all $\eta\in\Gamma_0$ and all $\gamma\in\Gamma$ 
by 
$$
E(\eta ):=\sum_{\{x,y\}\subset \eta }\phi (x,y),\quad 
E(\emptyset):=E(\{x\}):=0 
$$
and
\begin{eqnarray*}
W(\eta ,\gamma ):=\left\{ 
\begin{array}{cl}
\displaystyle\sum_{x\in \eta ,y\in \gamma }\phi (x,y), & \mathrm{if\;}
\displaystyle\sum_{x\in \eta ,y\in \gamma }|\phi (x,y)|<\infty \\ 
&  \\ 
+\infty , & \mathrm{otherwise}
\end{array}
\right. ,
\end{eqnarray*}
respectively. We set $W(\emptyset,\gamma ):= W(\eta ,\emptyset):= 0$. A grand 
canonical Gibbs measure (Gibbs measure for short) corresponding to a pair 
potential $\phi$, the intensity measure $\sigma$, and an inverse temperature 
$\beta>0$, is usually defined through the Dobrushin-Lanford-Ruelle equation. 
For convenience, we present here an equivalent definition through the 
Georgii-Nguyen-Zessin equation ((GNZ)-equation) (\cite[Theorem 2]{NZ79}, see 
also \cite[Theorem 3.12]{KoKu98}, \cite[Appendix A.1]{K00}). More precisely, 
a probability measure $\mu$ on $(\Gamma, \mathcal{B}(\Gamma))$ is called a 
Gibbs measure if it fulfills the integral equation
\begin{equation}
\int_\Gamma \sum_{x\in \gamma }H(x,\gamma\!\setminus\!\{x\})\,d\mu (\gamma )
=\int_\Gamma \int_XH(x,\gamma )e^{-\beta W(\{x\},\gamma )}\,
d\sigma(x)d\mu (\gamma ) \label{1.3}
\end{equation}
for all positive measurable functions $H:X\times\Gamma\to \R$. In 
particular, for $\phi\equiv 0$, (\ref{1.3}) reduces to the Mecke identity, 
which yields an equivalent definition of the Poisson measure $\pi_\sigma$ 
\cite[Theorem 3.1]{Me67}. 

Correlation measures corresponding to Gibbs measures are always absolutely 
continuous with respect to the Lebesgue-Poisson measure $\lambda_\sigma$. In 
view of this fact and Remark \ref{M.4}, the framework used throughout this
section is restricted to measures 
$\mu \in \mathcal{M}_{\mathrm{fexp}}^1(\Gamma )$ that are locally absolutely 
continuous with respect to the Poisson measure $\pi _\sigma$. Furthermore,  
we shall assume that the corresponding correlation functions $k_\mu$ fulfill 
the so-called Ruelle type bound inequality, that is, there are a $a>0$ and a 
$0<\varepsilon\leq 1$ such that 
\[
k_\mu(\eta )\leq \left( \left| \eta \right| !\right) ^{1-\varepsilon
}e_\lambda(a,\eta )=\left( \left| \eta \right| !\right)
^{1-\varepsilon }a^{\left| \eta \right| },\quad \lambda _{_{}\sigma }\mathrm{%
-a.a.}\text{ }\eta \in \Gamma _0. 
\]
According to Proposition \ref{9Prop9.1.10}, this assumption implies that
\begin{enumerate}
\item[1.] There are $c_1,c_2>0$ such that 
\[
\left|L_\mu(\theta)\right|\leq 
c_1\exp\left(c_2 \|\theta\|^{1/\varepsilon}_{L^1(\sigma)}\right)
\quad \text{for all } \theta \in L^1(\sigma).
\]
\end{enumerate}

As a consequence of Proposition \ref{9Prop9.1.10}, the Bogoliubov functional 
$L_\mu$ is entire of bounded type on $L^1(\sigma)$.

To proceed towards the equivalent description of Gibbs measures through 
Bogoliubov functionals, we consider potentials $\phi$ fulfilling 
the following semi-boundedness and integrability conditions:
\begin{enumerate}
\item[2.] $\exists B\geq 0:\quad \phi (x,y)\geq -2B\quad \mathrm{for\,\,all\,\,}x,y\in X$
\item[3.] $C(\beta ):=\displaystyle\esssup_{x\in X}\int_X\left|
e^{-\beta \phi (x,y)}-1\right| d\sigma (y)<\infty$
\end{enumerate}

\begin{proposition}
\label{9Prop9.2.2}Given a $\mu\in \mathcal{M}_{\mathrm{fexp}}^1(\Gamma )$ and 
a pair potential $\phi$, assume that Assumptions 1--3 are fulfilled. Then 
$\mu$ is a Gibbs measure corresponding to the potential $\phi $, the intensity
measure $\sigma $, and the inverse temperature $\beta $ if and only if the
Bogoliubov functional $L_\mu$ corresponding to $\mu $ solves the so-called
Bogoliubov (equilibrium) equation,
\[
\frac{\delta L(\theta )}{\delta \theta (x)}=L\left( (1+\theta )\left(
e^{-\beta \phi (x,\cdot )}-1\right) +\theta \right),\quad 
\sigma \mathrm{-\mathit{a.e.}}, 
\]
for all $\theta \in L^1(\sigma )$.
\end{proposition}

\noindent
\textbf{Proof.} 
The analyticity of $L_\mu$ on $L^1(\sigma )$ implies 
\begin{eqnarray}
dL_\mu(\theta ;f)&=&
\int_{\Gamma}\frac{d}{dz}\prod_{x\in\gamma}(1+\theta(x) +zf(x))
\Big\vert_{z=0}\,d\mu (\gamma)\nonumber \\
&=&\int_\Gamma \sum_{x\in \gamma }f(x)\prod_{y\in
\gamma \backslash \{x\}}(1+\theta (y))\,d\mu (\gamma ),\quad \theta ,f\in
L^1(\sigma ).\label{DuarteII} 
\end{eqnarray}
Thus, for a Gibbs measure $\mu$, the (GNZ)-equation yields for the right-hand 
side of (\ref{DuarteII})
\begin{equation}
\int_X\!f(x)\!\int_\Gamma \prod_{y\in \gamma
}(1+\theta (y))e^{-\beta W(\{x\},\gamma )}\,d\mu (\gamma )d\sigma (x).\label{DuarteIII}
\end{equation}
We claim that
\begin{equation}
e^{-\beta W(\{x\},\gamma )}=\prod_{y\in\gamma}
\left( 1+\left( e^{-\beta \phi (x,y)}-1\right)\right),\label{DuarteIV}
\end{equation}
which proof we postpone to the end. Hence (\ref{DuarteIII}) is given by
\begin{eqnarray*}
&&\int_X\!f(x)\!\int_\Gamma \prod_{y\in \gamma }(1+\theta (y))\prod_{y\in
\gamma }\left( 1+\left( e^{-\beta \phi (x,y)}-1\right) \right) d\mu (\gamma
)d\sigma (x)\\
&=&\!\!\int_X\!f(x)\!\int_\Gamma \prod_{y\in \gamma }
\left( (1+\theta (y))\left( e^{-\beta \phi (x,y)}-1\right) + 1 + \theta(y)\right) d\mu (\gamma)d\sigma (x).
\end{eqnarray*}
In this way we show that for all $f\in L^1(\sigma)$
\[
dL_\mu(\theta ;f)= \int_X f(x) 
L_\mu\left((1+\theta)(e^{-\beta \phi(x,\cdot)}-1)+\theta\right)\,d\sigma(x), 
\]
provided $(1+\theta)(e^{-\beta \phi(x,\cdot)}-1)+\theta\in L^1(X,\sigma)$.
Assumption 1 then implies that 
$L_\mu\left((1+\theta)(e^{-\beta \phi(x,\cdot)}-1)+\theta\right)\in L^\infty(X,\sigma)$ which completes the first part of the proof. Conversely, the same 
arguments as before yield,
\begin{eqnarray*}
&&\int_\Gamma \sum_{x\in \gamma }f(x)\prod_{y\in \gamma \backslash
\{x\}}(1+\theta (y))\,d\mu (\gamma )=dL_\mu(\theta ;f) \\
&=&\int_Xf(x)L_\mu\left( (1+\theta )\left( e^{-\beta \phi (x,\cdot
)}-1\right) +\theta \right)\,d\sigma (x) \\
&=&\int_Xf(x)\int_\Gamma \prod_{y\in \gamma }(1+\theta (y))e^{-\beta
W(\{x\},\gamma )}d\mu (\gamma )\,d\sigma (x),
\end{eqnarray*}
showing that the measure $\mu$ fulfills the (GNZ)-equation for the class of 
functions $H$ of the form
\[
H(x,\gamma )=f(x)\prod_{y\in \gamma \backslash \{x\}}(1+\theta (y)),\quad
\theta ,f\in L^1(\sigma ). 
\]
The result follows by a monotone class argument.

To conclude this proof amounts to check the technical problems left open. Due 
to Assumptions 2 and 3 one has
$$
\left\|\theta e^{-\beta \phi (x,\cdot)}
+e^{-\beta \phi (x,\cdot)}-1\right\|_{L^1(\sigma)}
\leq e^{2\beta B}\|\theta\|_{L^1(\sigma)} + C(\beta),
$$
showing $(1+\theta)(e^{-\beta \phi(x,\cdot)}-1)+\theta\in L^1(X,\sigma)$.

The infinite product 
$\prod_{y\in\gamma}(1+| e^{-\beta \phi (x,y)}-1|)$ converges for 
$\sigma\otimes\mu$-a.a.~$(x,\gamma)$, because Assumption 3 implies that 
$\sigma$-a.e.~$\Vert e^{-\beta \phi (x,\cdot)}-1\Vert_{L^1(\sigma)}<\infty$ and
\[
\int_\Gamma\prod_{y\in\gamma}\left( 1+\left |e^{-\beta\phi (x,y)}-1\right |\right)\,d\mu(\gamma)<\infty.
\]
The absolute convergence of the infinite product in (\ref{DuarteIV}) implies 
the convergence of $\sum_{y\in\gamma}| e^{-\beta \phi (x,y)}-1|$. Hence, 
either the series $\sum_{y\in\gamma}|\phi (x,y)|$ converges or there is a 
$y\in\gamma$ such that $\phi(x,y)=+\infty$. In the latter case the infinite 
product in (\ref{DuarteIV}) as well as $e^{-\beta W(\{x\},\gamma)}$ are both 
zero. For the first case we obtain
\[
\prod_{y\in\gamma}\left( 1+\left( e^{-\beta \phi (x,y)}-1\right)\right)
= \exp\left(-\beta\sum_{y\in\gamma}\phi(x,y)\right)
=e^{-\beta W(\{x\},\gamma)}.
\]   
\hfill $\blacksquare \medskip$

For higher order derivatives the corresponding Bogoliubov equations are 
defined as follows.

\begin{corollary}
\label{9Prop9.2.3}Given a $\mu\in \mathcal{M}_{\mathrm{fexp}}^1(\Gamma )$ and 
a pair potential $\phi$, assume that Assumptions 1--3 are fulfilled. If $\mu$ 
is a Gibbs measure corresponding to the potential $\phi $, the intensity 
measure $\sigma $, and the inverse temperature $\beta $, then for all 
$\theta \in L^1(\sigma )$ the following relation holds: 
\[
\left(D^n L_\mu\right)(\theta;\eta)=e^{-\beta
E(\eta )}L_\mu\!\left( (1+\theta )\left( e^{-\beta W(\eta ,\{\cdot
\})}-1\right) +\theta \right), \quad \sigma ^{(n)}\mathrm{-\mathit{a.a.}\,\,}\eta \in \Gamma _X^{(n)}. 
\]
\end{corollary}

\noindent
\textbf{Proof.} It follows from successive applications of Proposition 
\ref{9Prop9.2.2} and the chain rule to the function 
$L^1(\sigma )\ni \theta \mapsto (1+\theta
)\left( e^{-\beta \phi (x,\cdot )}-1\right) +\theta \in L^1(\sigma )$.\hfill 
$\blacksquare \medskip$

\begin{proposition}
\label{9Prop9.2.4}For any pair potential $\phi$ and any measure 
$\mu\in\mathcal{M}_{\mathrm{fexp}}^1(\Gamma )$ under Assumptions 1--3, the 
following equations are e\-quiv\-a\-lent:\newline
\smallskip \noindent (i) For all $\theta \in L^1(\sigma )$, 
\[
\frac{\delta L_\mu(\theta )}{\delta \theta (x)}=L_\mu\left(
(1+\theta )\left( e^{-\beta \phi (x,\cdot )}-1\right) +\theta \right) \quad 
\mathit{for}\mathrm{\,\,}\sigma \mathrm{-\mathit{a.a.}\,\,}x\in X. 
\]
\noindent (ii) For every $\theta ,f\in L^1(\sigma )$, 
\begin{eqnarray*}
&&L_\mu(\theta +f)-L_\mu(\theta ) \\
&=&\int_Xf(x)\int_0^1L_\mu\left( (1+\theta +tf)\left( e^{-\beta \phi
(x,\cdot )}-1\right) +\theta +tf\right) dtd\sigma (x).
\end{eqnarray*}
Furthermore, the previous equations imply that\newline
\smallskip \noindent (iii) For all $\theta ,f\in L^1(\sigma )$, 
\[
L_\mu(\theta +f)=\int_{\Gamma _0}e_\lambda(f,\eta )e^{-\beta
E(\eta )}L_\mu\left( (1+\theta )\left( e^{-\beta W(\eta ,\{\cdot
\})}-1\right) +\theta \right) d\lambda _\sigma (\eta ). 
\]
\end{proposition}

\begin{remark}
Assumptions 1--3 are not sufficient to insure the existence of the integral on
the right-hand side of the equation stated in (iii).
\end{remark}

\noindent 
\textbf{Proof.} $\mathrm{(i)}\Rightarrow \mathrm{(ii)}$: Since $L_\mu$ is 
entire on $L^1(\sigma )$, one has 
\[
L_\mu(\theta +f)-L_\mu(\theta )=\int_0^1\frac d{dt}L_\mu
(\theta +tf) dt 
\]
and, according to (i),
\begin{eqnarray*}
&&\frac d{dt}L_\mu(\theta +tf) =dL_\mu(\theta+tf,f)\\
&=&\int_Xf(x)L_\mu\left( (1+\theta +tf)\left( e^{-\beta \phi (x,\cdot
)}-1\right) +\theta +tf\right) d\sigma (x).
\end{eqnarray*}

\noindent
$\mathrm{(ii)}\Rightarrow \mathrm{(i)}$: Assuming (ii), for any 
$\theta ,f\in L^1(\sigma )$ one finds
\begin{eqnarray*}
\lefteqn{\frac d{dz}L_\mu(\theta +zf)\Big\vert_{z=0} 
=\lim_{z\rightarrow 0}\frac{L_\mu(\theta +zf)-L_\mu(\theta )}z }\\
&=&\lim_{z\rightarrow 0}\int_Xf(x)\int_0^1L_\mu\left( (1+\theta
+tzf)\left( e^{-\beta \phi (x,\cdot )}-1\right) +\theta +tzf\right)
dtd\sigma (x).
\end{eqnarray*}
Assumptions 1--3 allow to apply the Lebesgue dominated convergence
theorem and thus, interchanging the limit with the integrals and using the
continuity of $L_\mu$ on $L^1(\sigma )$, to obtain 
\[
\int_Xf(x)L_\mu\left( (1+\theta )\left( e^{-\beta \phi (x,\cdot
)}-1\right) +\theta \right) d\sigma (x). 
\]

\noindent
$\mathrm{(i)}\Rightarrow \mathrm{(iii)}$: The analyticity of 
$L_\mu$ straightforwardly leads (Remark \ref{Verdasca}) to
\begin{eqnarray*}
L_\mu(\theta +f) &=&\int_{\Gamma _0}e_\lambda(f,\eta )
\left(D^{\left|\eta\right|} L_\mu\right)(\theta;\eta)
d\lambda _\sigma (\eta ) \\
&=&\int_{\Gamma _0}e_\lambda(f,\eta )e^{-\beta E(\eta )}L_\mu
\left( (1+\theta )\left( e^{-\beta W(\eta ,\{\cdot \})}-1\right) +\theta
\right) d\lambda_\sigma (\eta ),
\end{eqnarray*}
where the second equality is a consequence of Corollary~\ref{9Prop9.2.3}.
\hfill $\blacksquare \medskip$

Proposition \ref{9Prop9.2.4} leads to a uniqueness result for Gibbs measures
corresponding to positive potentials. As a first step towards this purpose, we 
must introduce additional spaces of functionals. More precisely, for each 
$\alpha >0$, let \textrm{Ent}$_\alpha(L^1(\sigma ))$ be the space of all 
entire functionals 
$L$ on $L^1(\sigma )$ such that 
\[
\left\| L\right\| _\alpha :=\sup_{\theta \in L^1(\sigma )}\left( \left|
L(\theta )\right| e^{-\alpha \left\| \theta \right\| _{L^1(\sigma )}}\right)
<\infty . 
\]
It is clear that $\left\| \cdot \right\| _\alpha$ defines a norm on 
\textrm{Ent}$_\alpha(L^1(\sigma ))$.

\begin{proposition}
\label{9Prop9.2.7}With respect to the norm $\left\| \cdot \right\|
_\alpha $, $\mathrm{Ent}_\alpha (L^1(\sigma ))$ has the structure
of a Banach space.
\end{proposition}

\noindent
\textbf{Proof.} Fixing an $\alpha >0$, let $(L_n)_{n\in \N}$ be a Cauchy 
sequence in \textrm{Ent}$_\alpha (L^1(\sigma ))$, i.e., $(L_ne^{-\alpha \left\|
\cdot \right\| _{L^1(\sigma )}})_{n\in \N}$ is a Cauchy sequence
in the Banach space consisting of all complex-valued bounded functions
defined on $L^1(\sigma )$ endowed with the supremum norm. By completeness,
there is a complex-valued bounded function $\bar{L}$ such that 
\begin{equation}
\lim_{n\rightarrow \infty }\sup_{\theta \in L^1(\sigma )}\left( \left|
L_n(\theta )e^{-\alpha \left\| \theta \right\| _{L^1(\sigma )}}-\bar{L}
(\theta )\right| \right) =0.  \label{9Eq2.3}
\end{equation}
It remains to show that the functional 
$L(\theta ):=\bar{L}(\theta)e^{\alpha \left\| \theta \right\| _{L^1(\sigma )}}$, $\theta \in L^1(\sigma )$, is entire on $L^1(\sigma )$. This 
follows from the Vitali theorem (see e.g.~\cite{HP57}), since by 
(\ref{9Eq2.3}) the sequence $(L_n)_{n\in \N}$ converges pointwisely to $L$ 
and, by the inequality 
\begin{eqnarray*}
\left| L_n(\theta )\right| &\leq &\sup_{\theta \in L^1(\sigma )}\left(
\left| L_n(\theta )\right| e^{-\alpha \left\| \theta \right\| _{L^1(\sigma
)}}\right) e^{\alpha \left\| \theta \right\| _{L^1(\sigma )}} \\
&=&\left\| L_n\right\| _\alpha e^{\alpha \left\| \theta \right\|
_{L^1(\sigma )}},\quad n\in \N,
\end{eqnarray*}
the sequence $(L_n)_{n\in\N}$ is locally uniformly bounded in 
$L^1(\sigma )$.\hfill $\blacksquare \medskip$

For pair potentials $\phi$ semi-bounded from below fulfilling Assumption~3, 
Proposition \ref{9Prop9.2.4} has shown that any functional $L$ in 
$\mathrm{Ent}_\alpha (L^1(\sigma ))$ solving the initial value problem 
\[
\left\{ 
\begin{array}{l}
\displaystyle\frac{\delta L(\theta )}{\delta \theta (x)}=L\left(
(1+\theta )\left( e^{-\beta \phi (x,\cdot )}-1\right) +\theta \right) ,\quad
\theta \in L^1(\sigma ) \\ 
\\ 
L(0)=1
\end{array}
\right. 
\]
is a solution of the equation 
\[
L(\theta )-1=\int_X\theta (x)\int_0^1L\left( (1+t\theta )\left( e^{-\beta
\phi (x,\cdot )}-1\right) +t\theta \right) dtd\sigma (x),\quad \theta \in
L^1(\sigma ). 
\]
In the sequel we denote by $J$ the linear mapping defined on each 
space \textrm{Ent}$_\alpha (L^1(\sigma ))$, $\alpha > 0$, by 
\[
\left( JL\right) (\theta ):=\int_X\theta (x)\int_0^1L\left( (1+t\theta
)\left( e^{-\beta \phi (x,\cdot )}-1\right) +t\theta \right) dtd\sigma (x), 
\]
for
 $L\in \mathrm{Ent}_\alpha (L^1(\sigma )),\theta \in L^1(\sigma )$.

\begin{proposition}
\label{9Prop9.2.5}Let $\phi $ be a positive pair potential fulfilling 
Assumption 3. Then, for any $\alpha >0$, the mapping $J$ defines a bounded 
linear operator on $\mathrm{Ent}_\alpha (L^1(\sigma ))$. Moreover, for all 
$L\in \mathrm{Ent}_\alpha (L^1(\sigma ))$,
\[
\left\| JL\right\| _\alpha \leq \frac{e^{\alpha C(\beta )}}\alpha
\left\| L\right\| _\alpha .
\]
\end{proposition}

\noindent
\textbf{Proof.} Let $\alpha >0$ be given. For all $\theta \in L^1(\sigma )$ 
one has 
\[
\left| \left( JL\right) (\theta )\right| \leq \left\| L\right\| _{_{}\alpha
}\int_X\left| \theta (x)\right| \int_0^1e^{\alpha \left\| (1+t\theta )\left(
e^{-\beta \phi (x,\cdot )}-1\right) +t\theta \right\| _{L^1(\sigma
)}}dtd\sigma (x) 
\]
and, according to the stated assumptions on $\phi$, 
\begin{eqnarray*}
&&\left\| (1+t\theta )\left( e^{-\beta \phi (x,\cdot )}-1\right) +t\theta
\right\| _{L^1(\sigma )} \\
&\leq &t\int_X\left| \theta (y)\right| e^{-\beta \phi (x,y)}d\sigma
(y)+\int_X\left| e^{-\beta \phi (x,y)}-1\right| d\sigma (y) \\
&\leq &t\left\| \theta \right\| _{L^1(\sigma )}+C(\beta ).
\end{eqnarray*}
Therefore
\begin{eqnarray*}
\left| \left( JL\right) (\theta )\right| &\leq &\left\| L\right\|
_\alpha \left\| \theta \right\| _{L^1(\sigma )}e^{\alpha C(\beta
)}\int_0^1e^{\alpha t\left\| \theta \right\| _{L^1(\sigma )}}dt \\
&=&\left\| L\right\| _\alpha \frac{e^{\alpha C(\beta )}}\alpha \left(
e^{\alpha \left\| \theta \right\| _{L^1(\sigma )}}-1\right) <\left\|
L\right\| _\alpha \frac{e^{\alpha C(\beta )}}\alpha e^{\alpha \left\|
\theta \right\| _{L^1(\sigma )}},
\end{eqnarray*}
showing the required estimate of the norms.\hfill $\blacksquare \medskip$

\begin{corollary}
\label{9Prop9.2.6}Let $\beta >0$ be given. Then, under the assumptions of
Proposition \ref{9Prop9.2.5}, on each space $\mathrm{Ent}_\alpha(L^1(\sigma))$ 
with 
\[
\frac{e^{\alpha C(\beta )}}\alpha <1 
\]
exists a unique solution of the equation 
\begin{equation}
L-JL=1.  \label{9Eq2.4}
\end{equation}
In particular, for all $\beta >0$ such that $C(\beta )<e^{-1}$, there is a
unique solution of equation (\ref{9Eq2.4}) for a suitable choice of $\alpha $
(\textit{e.g}., $\alpha =(C(\beta ))^{-1}$).
\end{corollary}

\noindent
\textbf{Proof.} According to Proposition \ref{9Prop9.2.5}, one has 
\[
\left\| JL\right\| _\alpha \leq \frac{e^{\alpha C(\beta )}}\alpha
\left\| L\right\| _\alpha <\left\| L\right\| _\alpha,\quad L\in 
\mathrm{Ent}_\alpha (L^1(\sigma )).
\]
That is, the operator $J$ is a contraction on 
$\mathrm{Ent}_\alpha(L^1(\sigma ))$. Thus, by the contraction mapping 
principle, there is a unique solution of equation (\ref{9Eq2.4}), namely, 
$(1-J)^{-1}1$, with $(1-J)^{-1}$ defined by the von Neumann 
series $\sum_{n=0}^\infty J^n$. The last assertion follows by minimizing the
expression $\alpha ^{-1}e^{\alpha C(\beta )}$ in the parameter $\alpha $.
\hfill $\blacksquare \medskip$

In this way we have proved the following uniqueness result.

\begin{theorem}
\label{9Prop9.2.8}Let $\phi $ be a positive pair potential fulfilling
the integrability condition 
\[
C(\beta )=\esssup_{x\in X}\int_X\left| e^{-\beta
\phi (x,y)}-1\right| d\sigma (y)<\infty . 
\]
For each $\beta >0$ such that $C(\beta )<e^{-1}$ there is at most one Gibbs
measure fulfilling Ruelle bound and corresponding to the potential 
$\phi $, the intensity measure $\sigma$, and the inverse temperature 
$\beta $. 
\end{theorem}

\section{Stochastic dynamic equations\label{Section5}}

To deal with the differential structures used below to study a diffusion 
dynamics of a continuous system, this section begins by recalling a few
concepts of the intrinsic geometry on configuration spaces (\cite{AKR97}, 
\cite{KoKu99}, \cite{K00}).  

\subsection{Differential geometry on configuration spaces}

Apart from the topological structure, the bijection defined in Section 
\ref{Section2} between the spaces $\Gamma _X^{(n)}$ and 
$\widetilde{X^n}/S_n$ also induces a differentiable structure on 
$\Gamma _X^{(n)}$ (see (\ref{1Eq1.2})). More precisely, given $n$ charts 
$(h_1,U_1),...,(h_n,U_n)$ of $X$, where $U_1,...,U_n$ are mutually disjoint 
open sets in $X$, one constructs a chart $h_1\hat{\times}...\hat{\times}h_n$ 
of $\Gamma _X^{(n)}$ defined on the open set 
$U_1\hat{\times}...\hat{\times}U_n$ in $\Gamma _X^{(n)}$, 
\[
U_1\hat{\times}...\hat{\times}U_n:=\left\{ \eta =\{x_1,...,x_n\}\in \Gamma
_X^{(n)}:\exists \iota \in S_n\mathrm{\ s.t.}\text{ }x_{\iota (k)}\in
U_k,k=1,...,n\right\} , 
\]
by 
\[
\left( h_1\hat{\times}...\hat{\times}h_n\right) \left(
\{x_1,...,x_n\}\right) :=\left( h_1(x_{\iota (1)}),...,h_n(x_{\iota
(n)})\right)\!\in\!h_1(U_1)\times ...\times h_n(U_n). 
\]
Each set $\Gamma _X^{(n)}$ endowed with this geometry has the structure of a 
$n\cdot \mathrm{dim}(X)$-dimensional $C^\infty $-manifold. In this way we 
have also defined a differentiable structure on $\Gamma _0$. For any vector 
field $v$ on $X$ we have
\[
\left( \nabla _v^{\Gamma _0}G\right) (\eta )=\sum_{x\in \eta }\left\langle
\left( \nabla ^{\Gamma _0}G\right) (\eta ,x),v(x)\right\rangle _{T_xX},
\]
yielding, in particular,
\begin{equation}
\left( \nabla ^{\Gamma _0}e_\lambda(\theta )\right) (\eta ,x)=\nabla
^X\theta (x)e_\lambda(\theta ,\eta \backslash \{x\}),\quad \eta \in
\Gamma _0,x\in \eta ,  \label{9Eq3.1}
\end{equation}
$\nabla :=\nabla ^X$ being the gradient on $X$. For the Laplace-Beltrami
operator $\triangle ^{\Gamma _0}$ on $\Gamma _0$, which is defined by the
direct sum of the Laplace-Beltrami operators $\triangle ^{\Gamma _X^{(n)}}$
on $\Gamma _X^{(n)}$, we find
\begin{equation}
\left( \triangle ^{\Gamma _0}e_\lambda(\theta )\right) (\eta
)=\sum_{x\in \eta }\triangle ^X\theta (x)e_\lambda(\theta ,\eta
\backslash \{x\}),  \label{9Eq3.2}
\end{equation}
where $\triangle :=\triangle ^X$ denotes the Laplace-Beltrami operator on $X$.

In the sequel we use the classical notation $C^k(\Gamma _0)$, 
$k\in\N\cup\{\infty\}$, for the space of all real-valued $C^k$-functions on 
$\Gamma_0$, and $C_{0}^k(\Gamma _0)$ for the space of all functions $G$ in
$C^k(\Gamma _0)$ with bounded support such that 
for some $\varepsilon >0$ one has $G(\eta)=0$ for all $\eta$ 
containing a pair $x,y$, $x\neq y$,
such that $|x-y|\leq \varepsilon$. 

Through the $K$-transform one may introduce a differential structure on 
$\Gamma$ \cite{KoKu99}, which coincides with the one introduced in 
\cite{AKR97} by ''lifting'' the geometrical structure on the underlying 
manifold $X$. For each $G\in C_{0}^1(\Gamma _0)$, 
\[
\left( \nabla ^\Gamma (KG)\right) (\gamma ,x):=
\sum_{\stackrel{\scriptstyle \eta \subset \gamma:\,\vert\eta\vert<\infty, }{x\in \eta }}
\left( \nabla ^{\Gamma _0}G\right) (\eta,x),\quad 
\gamma \in \Gamma ,x\in \gamma , 
\]
and $\triangle ^\Gamma :=K\triangle ^{\Gamma _0}K^{-1}$ on 
$\mathcal{FP}(C_{0}^2,\Gamma)$, the set of all twice differentiable
cylinder polynomials $F$ with the property that there exists 
a $\varepsilon >0$ such that $F(\gamma)=0$ on all $\gamma$ which contains a 
pair of points in the domain of cylindricity with distance smaller than 
$\varepsilon$. Equivalently, all such functions $F$ are of the form 
$F=KG$, $G\in C_{0}^2(\Gamma _0)$.

\subsection{Non-equilibrium stochastic dynamics equations\label{Subsection5.2}}

The purpose of this subsection is to investigate the problem 
heuristically formulated in (\ref{9Eq3.5}). Let us first fix the framework. 
On the space $X=\R^d$, $d\in\N$, let us consider the intensity measure 
$d\sigma(x)=zdm(x)$, $m$ being the Lebesgue measure on $\R^d$ and $z>0$
(activity), 
and a meas\-urable function $V:\R^d\to\R\cup\{+\infty\}$ (potential)
such that $V(-x)=V(x)\in\R$ for all $x\in\R^d\setminus\{0\}$. Accordingly,
we may define a translation invariant pair potential $\phi$ on $\R^d$ by 
$\phi (x,y):=V(y-x)$. Concerning $V$, we must at least assume the
standard Ruelle conditions of superstability, integrability (i.e., 
Assumption 3), and lower regularity (\cite{Ru70}), which are sufficient to 
insure the
existence of corresponding Gibbs measures, cf.~e.g.~\cite[Section 5]{Ru70}. 
In particular, this includes the class of potentials
$V$ which are bounded from below and
integrable at infinity, and having a small enough negative part. 

The problem under consideration is the construction of a solution to the 
system of stochastic differential equations heuristically given by
\begin{equation}
\left\{ 
\begin{array}{l}
\displaystyle dx_k(t)=-\frac\beta 2\sum_{1\leq i\neq k}\nabla V(x_k(t)-x_i(t))dt+dW_k(t),\quad t\geq 0 \\ 
\\ 
x_k(0)=x_k,\quad k\in \N
\end{array}
\right. ,\label{Ku051}
\end{equation}
where $W_k$, $k \in \mathbb{N}$, is a family of independent Brownian 
motions. Note that due to the symmetry in the labels, any solution 
$(x_k)_k$ in $({\mathbb{R}}^d)^{\mathbb{N}}$ of (\ref{Ku051}) can be 
interpreted (modulo collapse) as a stochastic process with paths in 
configuration space, that is,
$\gamma(t):=\{x_k(t): k \in \mathbb{N}\}$. Informally, the generator of this 
dynamics is given by
\[
\left( HF\right) (\gamma ):=-\frac 12\left( \triangle ^\Gamma F\right)
(\gamma )+\frac \beta 2\!\sum_{x\in\gamma}\sum_{y\in\gamma\setminus\{x\}}
\!\!\left\langle\nabla _xV(x-y),\nabla ^\Gamma F(\gamma ,x)\right\rangle ,  
\]
where $\left\langle \cdot ,\cdot\right\rangle$ denotes the inner product on 
$\R^d$ and $\beta$ the inverse temperature. Note that in contrast to
(\ref{Ku051}), the generator $H$ is well-defined, for example, on 
$\mathcal{FP}(C_{0}^2,\Gamma)$.

In the equilibrium dynamics case, the authors in \cite{AKR97a} have 
constructed a solution for a wide class of potentials $V$. More precisely, for a Gibbs measure $\mu_\mathrm{inv}$ 
corresponding to $V$, the same as used in definition (\ref{Ku051}), it has 
been shown that $H$ is a positive symmetric operator on the 
space $L^2(\Gamma,\mu_\mathrm{inv})$ associated to the
Dirichlet form 
\[
\left( HF,F\right) _{L^2(\mu_\mathrm{inv})}=\frac 12\int_\Gamma \sum_{x\in \gamma
}\left| \nabla ^\Gamma F(\gamma ,x)\right| ^2d\mu_\mathrm{inv} (\gamma ). 
\]
This allows the use of standard Dirichlet form techniques to construct a 
diffusion process corresponding to $H$ having $\mu_\mathrm{inv}$ as an 
invariant (and, moreover, reversible) measure and starting on 
$\mu_\mathrm{inv}$-a.a.~initial points. This yields, in particular, 
the corresponding semigroup $T_t := e^{-Ht}$, $t\geq 0$, on 
$L^2(\Gamma,\mu_\mathrm{inv})$, a solution of the Cauchy problem
\[
\left\{ 
\begin{array}{l}
\displaystyle\frac d{dt}F_t=-HF_t,\quad t\geq 0 \\ 
\\ 
F_0
\end{array}
\right. .  
\]
For further references see also \cite{AKR97a}.

An essentially more difficult and interesting question is the 
non-equilibrium dynamics case. This means, the construction of the dynamics 
without reference to any invariant measure. In this case, the above scheme 
does not apply, and the only general result was obtained by \cite{Fr87} for a 
restrictive class of potentials and $d\leq 4$.

In the sequel we describe a new scheme for the construction of the dynamics, 
based on the diagram in Remark \ref{Rem1} (Section \ref{Section2}). For this 
purpose we shall fix a probability measure $\mu$ on $\Gamma$ as an initial 
distribution. In 
contrast to the previous situation, we now assume that the measure $\mu$ is 
neither an invariant measure nor a perturbation of an invariant one.

The starting point for the approach is the description of the operator $H$ in 
terms of quasi-observables. In fact, as $H$ is well-defined, for instance, on 
$\mathcal{FP}(C_{0}^2,\Gamma)$, its image under the $K$-transform yields on the space of 
quasi-observables the operator $\hat{H}:=K^{-1}HK$ acting on functions 
$G\in C_{0}^2(\Gamma _0)$ by 
\begin{eqnarray}
\left( \hat{H}G\right) (\eta ) &=&-\frac 12\left( \triangle ^{\Gamma
_0}G\right) (\eta )  \nonumber \\
&&+\frac \beta 2\sum_{x\in \eta }\sum_{y\in \eta \backslash \{x\}}\left\{
\left\langle \nabla _xV(x-y),\left( \nabla ^{\Gamma _0}G\right) (\eta
,x)\right\rangle \right.  \label{9Eq3.3} \\
&&+\left. \left\langle \nabla _xV(x-y),\left( \nabla ^{\Gamma _0}G\right)
(\eta \backslash \{y\},x)\right\rangle \right\} .  \nonumber
\end{eqnarray}

The time evolution equation is then given by the corresponding Cauchy problem
\begin{equation}
\left\{ 
\begin{array}{l}
\displaystyle\frac \partial {\partial t}G_t(\eta )=-\hat{H}G_t(\eta ),\quad
t\geq 0,\eta \in \Gamma _0 \\ 
\\ 
G_0\in C_{0}^\infty (\Gamma _0)
\end{array}
\right. , \label{Ku05dqo}
\end{equation}
having the advantage of being recursively solvable, because 
the time derivative of each $G_t\!\!\upharpoonright_{\Gamma^{(n)}_{\R^d}}$ 
depends only on $G_t\!\!\upharpoonright_{\Gamma^{(n)}_{\R^d}}$ and 
$G_t\!\!\upharpoonright_{\Gamma^{(n-1)}_{\R^d}}$. Hence, for 
quasi-observables, the evolution can be always constructed. However, the 
difficulty is to show that this solution is regular enough to allow a 
reconstruction of the dynamics on the level of functions on $\Gamma$. 

The previous procedure based on the diagram in Remark \ref{Rem1} allows to proceed 
further. Actually, we may also describe the dynamics in terms of correlation 
functions through the dual operator $\hat{H}^{*}$ of $\hat{H}$ in the sense 
\[
\int_{\Gamma_0}(\hat{H}G)(\eta) k(\eta)\,d\lambda_m(\eta)=
\int_{\Gamma_0}G(\eta) (\hat{H}^{*}k)(\eta)\,d\lambda_m(\eta).
\]
As an aside, let us mention that in the Hamiltonian dynamics case this 
approach corresponds to the well-known BBGKY-hierarchy, see e.g.~\cite{Bog46}. In our case, this leads to
\[
\left\{ 
\begin{array}{l}
\displaystyle\frac \partial {\partial t}k_t^{(n)}=-\left( \hat{H}%
^{*}k_t\right) ^{(n)} \\ 
\\ 
k_0^{(n)},\quad n\in\N_0
\end{array}
\right. ,  
\]
where $k_0^{(n)}$, $n\in\N_0$, are the correlation functions corresponding to 
the initial distribution $\mu$. This system of equations also has hierarchical 
structure in which the time derivative of each $k_t^{(n)}$ depends on 
$k_t^{(n+1)}$. Namely, written out explicitly, 
\begin{eqnarray*}
&&\frac \partial {\partial t}k_t^{(n)}(x_1,...,x_n)=\frac
12\sum_{k=1}^n\triangle _{x_k}k_t^{(n)}(x_1,...,x_n)  \\
&&+\frac \beta 2\sum_{\stackrel{\scriptstyle k,j=1}{k\neq j}}^n\triangle
V(x_k-x_j)k_t^{(n)}(x_1,...,x_n) \\
&&+\frac \beta 2\sum_{\stackrel{\scriptstyle k,j=1}{k\neq j}}^n\left\langle
\nabla _{x_k}V(x_k-x_j),\nabla _{x_k}k_t^{(n)}(x_1,...,x_n)\right\rangle \\
&&+\frac \beta 2\sum_{k=1}^n\int_{\R^d}\left\langle \nabla
_{x_k}V(x_k-y),\nabla _{x_k}k_t^{(n+1)}(x_1,...,x_n,y)\right\rangle dy \\
&&+\frac \beta 2\sum_{k=1}^n\int_{\R^d}\triangle
V(x_k-y)k_t^{(n+1)}(x_1,...,x_n,y)dy.  
\end{eqnarray*}
In theoretical physics this system of equations is known as the
Bogoliubov-Streltsova diffusion hierarchy (see \cite{S59}). 
We observe that the operator $\hat{H}^*$ can be rigorously defined, for 
example, on correlation functions $k$ fulfilling the bound 
\begin{equation}
|\Delta k(\eta )|+ |\nabla k(\eta )|+
k(\eta )\leq C^{\left| \eta \right| }e^{-\alpha E(\eta )},\quad
C,\alpha \geq 0,\ \lambda_m \mathrm{-a.a.}\text{ }\eta \in
\Gamma _0.  \label{9Eq3.11}
\end{equation}

Completing our way through the diagram (Remark \ref{Rem1}), one can construct 
a dynamics on states.

The previous construction implies, in particular, that the dynamics can 
be also expressed in terms of Bogoliubov functionals
\[
L_t(\theta ):=\int_{\Gamma _0}e_\lambda(\theta ,\eta )\rho_t(d\eta),
\quad t\geq 0. 
\]
This leads to the following result.

\begin{theorem}
\label{9Prop9.3.1}Under the above conditions one has
\begin{eqnarray}
\frac \partial {\partial t}L_t(\theta ) &=&\frac 12\int_{\R^d}\triangle
\theta (x)\frac{\delta L_t(\theta )}{\delta \theta (x)}dx  \nonumber \\
&&-\frac \beta 4\int_{\R^d}\int_{\R^d}\left\langle \nabla
_xV(x-y),\nabla \theta (x)(\theta (y)+1)\right.   \label{9Eq3.12} \\
&&\left. -\nabla \theta (y)(\theta (x)+1)\right\rangle \frac{\delta
^2L_t(\theta )}{\delta \theta (x)\delta \theta (y)}dxdy,  \nonumber
\end{eqnarray}
for all $\theta \in C_0^2(\R^d):=$ the space of all $C^2$-functions on $\R^d$ 
with compact support.
\end{theorem}

\begin{remark}
According to Theorem \ref{9Prop9.3.1}, the correlation functions $k_t$
corresponding to a solution $L_t$ for the diffusion hierarchical equation 
(\ref{9Eq3.12}) fulfill the generalized Ruelle bound (\ref{9Eq3.11}) 
\cite{KoKuKva}. 
\end{remark}

\noindent
\textbf{Proof.} According to (\ref{Ku05dqo}), 
\begin{eqnarray*}
\frac \partial {\partial t}L_t(\theta ) &=&\int_{\Gamma _0}e_{_{}\lambda
}(\theta ,\eta )\left( \frac d{dt}\rho _t\right) (d\eta ) \\
&=&-\int_{\Gamma _0}e_\lambda(\theta ,\eta )\left( \hat{H}^{*}\rho
_t\right) (d\eta ) \\
&=&-\int_{\Gamma _0}\left( \hat{H}e_\lambda(\theta )\right) (\eta
)\rho _t(d\eta )
\end{eqnarray*}
with 
\begin{eqnarray*}
\left( \hat{H}e_\lambda(\theta )\right) (\eta ) &=&-\frac 12\left(
\triangle ^{\Gamma _0}e_\lambda(\theta )\right) (\eta ) \\
&&+\frac \beta 2\sum_{x\in \eta }\sum_{y\in \eta \backslash \{x\}}\left\{
\left\langle \nabla _xV(x-y),\left( \nabla ^{\Gamma _0}e_\lambda
(\theta )\right) (\eta ,x)\right\rangle \right. \\
&&+\left. \left\langle \nabla _xV(x-y),\left( \nabla ^{\Gamma
_0}e_\lambda(\theta )\right) (\eta \backslash \{y\},x)\right\rangle
\right\} ,
\end{eqnarray*}
cf.~(\ref{9Eq3.3}). Therefore, equalities (\ref{9Eq3.1}) and (\ref{9Eq3.2})
yield 
\begin{eqnarray*}
&&\frac \partial {\partial t}L_t(\theta ) \\
&=&\frac 12\int_{\Gamma _0}\sum_{x\in \eta }\triangle \theta
(x)e_\lambda(\theta ,\eta \backslash \{x\})\rho _t(d\eta ) \\
&&-\frac \beta 2\int_{\Gamma _0}\sum_{x\in \eta }\sum_{y\in \eta \backslash
\{x\}}\left\langle \nabla _xV(x-y),\nabla \theta (x)\right\rangle
e_\lambda(\theta ,\eta \backslash \{x\})\rho _t(d\eta ) \\
&&-\frac \beta 2\int_{\Gamma _0}\sum_{x\in \eta }\sum_{y\in \eta \backslash
\{x\}}\left\langle \nabla _xV(x-y),\nabla \theta (x)\right\rangle
e_\lambda(\theta ,\eta \backslash \{x,y\})\rho _t(d\eta ) \\
&=&\frac 12\int_{\Gamma _0}\sum_{x\in \eta }\triangle \theta
(x)e_\lambda(\theta ,\eta \backslash \{x\})\rho _t(d\eta ) \\
&&-\frac \beta 2\int_{\Gamma _0}\sum_{\{x,y\}\subset \eta }\left\langle
\nabla _xV(x-y),\nabla \theta (x)\theta (y)-\nabla \theta (y)\theta
(x)\right\rangle \\
&&\cdot e_\lambda(\theta ,\eta \backslash \{x,y\})\rho _t(d\eta ) \\
&&-\frac \beta 2\int_{\Gamma _0}\sum_{\{x,y\}\subset \eta }\left\langle
\nabla _xV(x-y),\nabla \theta (x)-\nabla \theta (y)\right\rangle
e_\lambda(\theta ,\eta \backslash \{x,y\})\rho _t(d\eta ),
\end{eqnarray*}
and the proof follows by Corollary \ref{9Prop9.1.4}.\hfill 
$\blacksquare \medskip$

As a straightforward consequence, one may easily derive the time evolution
equation of the Laplace transform corresponding to the measures $\mu _t$, 
\[
\mathcal{L}_t(\varphi ):=\int_\Gamma \exp \left( \left\langle \gamma
,\varphi \right\rangle \right) \mu _t(d\gamma )=L_t\left( e^\varphi-1\right) . 
\]
In hydrodynamics this equation is related to the Hopf equation.

\begin{corollary}
\label{9Prop9.3.2}Under the conditions of Theorem \ref{9Prop9.3.1}, for
all $\varphi \!\in \!C_0^2(\R^d)$ we have 
\begin{eqnarray*}
\frac \partial {\partial t}\mathcal{L}_t(\varphi ) &=&\frac 12\int_{\R^d}
\left( \triangle \varphi (x)+\left| \nabla \varphi (x)\right| ^2\right) 
\frac{\delta \mathcal{L}_t(\varphi )}{\delta \varphi (x)}dx \\
&&-\frac \beta 4\int_{\R^d}\int_{\R^d}\left\langle \nabla
_xV(x-y),\nabla \varphi (x)-\nabla \varphi (y)\right\rangle  \\
&&\cdot \frac{\delta ^2\mathcal{L}_t(\varphi )}{\delta \varphi
(x)\delta \varphi (y)}dxdy.
\end{eqnarray*}
\end{corollary}

\noindent
\textbf{Proof.} In Theorem \ref{9Prop9.3.1} consider the case 
$\theta =e^\varphi -1$ with $\varphi \in C_0^2(\R^d)$. This gives 
\begin{eqnarray*}
\frac \partial {\partial t}\mathcal{L}_t(\varphi ) &=&\frac \partial
{\partial t}L_t(\theta ) \\
&=&\frac 12\int_{\R^d}\left( \triangle \varphi (x)+\left| \nabla
\varphi (x)\right| ^2\right) e^{\varphi (x)}\frac{\delta L_t(\theta )}
{\delta \theta (x)}dx \\
&&-\frac \beta 4\int_{\R^d}\int_{\R^d}\left\langle \nabla
_xV(x-y),\nabla \varphi (x)-\nabla \varphi (y)\right\rangle \\
&&\cdot e^{\varphi (x)+\varphi (y)}\frac{\delta ^2L_t(\theta )}
{\delta \theta (x)\delta \theta (y)}dxdy,
\end{eqnarray*}
and the proof follows because
\[
\frac{\delta \mathcal{L}_t(\varphi )}{\delta \varphi (x)}=\frac{\delta
L_t(\theta )}{\delta \theta (x)}\frac{\delta \left( e^\varphi-1\right) 
(\varphi )}{\delta \varphi (x)}=\frac{\delta L_t(\theta )}
{\delta \theta (x)}e^{\varphi (x)},\quad m\mathrm{-a.a.}\text{ }x 
\]
and 
\begin{eqnarray*}
\frac{\delta ^2\mathcal{L}_t(\varphi )}{\delta \varphi (x)\delta
\varphi (y)} &=&\frac \delta {\delta \varphi (x)}\left( \frac{\delta
\mathcal{L}_t(\varphi )}{\delta \varphi (y)}\right) =e^{\varphi
(y)}\frac \delta {\delta \varphi (x)}\left( \frac{\delta L_t(\theta )}
{\delta \theta (y)}\right) \\
&=&e^{\varphi (y)+\varphi (x)}\frac{\delta ^2L_t(\theta )}{\delta
\theta (x)\delta \theta (y)},\quad m\mathrm{-a.a.}\text{ }x,y.
\end{eqnarray*}
\hfill $\blacksquare $

\subsection*{Acknowledgments}

We truly thank Prof.~A.~Chebotarev for a careful reading of the manuscript and 
helpful comments. T.~K.~gratefully would like to acknowledge the support of 
the DAAD through a Postdoctoral fellowship, the A.~v.~Humboldt Foundation 
through a Feodor-Lynen fellowship, the DFG through
``Forschergruppe Spektrale Analysis, asymptotische Verteilungen
und stochastische Dynamik'', 
and NSF Grant (DMR 01-279-26). M.~J.~O.~is 
thankful to ''Sub\-programa Ci\^{e}ncia e Tecnologia do 2${{}^o}$ Quadro 
Comunit\'{a}rio de Apoio'' (PRAXIS XXI/BD/20000/99). This work was partially 
supported by FCT POCTI, FEDER.


\begin{thebibliography}{AKR98b}

\bibitem[AKR98a]{AKR97}
S.~Albeverio, {Yu}.~G. Kondratiev, and M.~R{\"o}ckner.
\newblock Analysis and geometry on configuration spaces.
\newblock {\em J.~Funct.~Anal.}, 154(2):444--500, 1998.

\bibitem[AKR98b]{AKR97a}
S.~Albeverio, {Yu}.~G. Kondratiev, and M.~R{\"o}ckner.
\newblock Analysis and geometry on configuration spaces: The {G}ibbsian case.
\newblock {\em J. Funct. Anal.}, 157:242--291, 1998.

\bibitem[Bar85]{Ba85}
J.~A. Barroso.
\newblock {\em Introduction to Holomorphy}, volume 106 of {\em Mathematics
  Studies}.
\newblock North-Holland Publ.~Co., Amsterdam, 1985.

\bibitem[Bog46]{Bog46}
N.~N. Bogoliubov.
\newblock {\em Problems of a Dynamical Theory in Statistical Physics}.
\newblock Gostekhisdat, Moscow, 1946.
\newblock (in Russian). English translation in J. de Boer and G. E. Uhlenbeck
  (editors), {\it Studies in Statistical Mechanics\/}, volume 1, pages 1--118.
  North-Holland, Amsterdam, 1962.

\bibitem[Din81]{Di81}
S.~Dineen.
\newblock {\em Complex Analysis in Locally Convex Spaces}, volume~57 of {\em
  Mathematics Studies}.
\newblock North-Holland Publ.~Co., Amsterdam, 1981.

\bibitem[DSI75]{DSI75}
M.~Duneau, B.~Souillard, and D.~Iagolnitzer.
\newblock Decay of correlations for infinite-range interactions.
\newblock {\em J. Math. Phys.}, 16(8):1662--1666, 1975.

\bibitem[DU77]{DiU79}
J.~Diestel and J.~J. Uhl.
\newblock {\em Vector Measures}, volume~15 of {\em Mathematical Surveys}.
\newblock Amer. Math. Soc., Providence, Rhode Island, 1977.

\bibitem[DVJ88]{DaVJ88}
D.~J. Daley and D.~Vere-Jones.
\newblock {\em An Introduction to the Theory of Point Processes}.
\newblock Springer Verlag, NewYork, Berlin, and Heidelberg, 1988.

\bibitem[FF91]{FicFre91}
K.-H. Fichtner and W.~Freudenberg.
\newblock Characterization of states of infinite {B}oson systems {I}. {On} the
  construction of states of {B}oson systems.
\newblock {\em Comm.~Math.~Phys.}, 137:315--357, 1991.

\bibitem[Fri87]{Fr87}
J.~Fritz.
\newblock Gradient dynamics of infinite point systems.
\newblock {\em Ann. Probab.}, 15(2):478--514, 1987.

\bibitem[HP57]{HP57}
E.~Hille and R.~S. Phillips.
\newblock {\em Functional Analysis and Semi-groups}, volume~31 of {\em
  Amer.~Math.~Soc.~Colloq.~Publ}.
\newblock American Mathematical Society, 1957.

\bibitem[KK02]{KoKu99}
{Yu}.~G. Kondratiev and T.~Kuna.
\newblock Harmonic analysis on configuration space {I}. {G}eneral theory.
\newblock {\em Infin.~Dimens.~Anal.~Quantum Probab.~Relat.~Top.},
  5(2):201--233, 2002.

\bibitem[KK03]{KoKu98}
{Yu}.~G. Kondratiev and T.~Kuna.
\newblock Correlation functionals for {G}ibbs measures and {R}uelle bounds.
\newblock {\em Methods Funct.~Anal.~Topology}, 9(1):9--58, 2003.

\bibitem[KK05]{KoKu99c}
{Yu}.~G. Kondratiev and T.~Kuna.
\newblock Harmonic analysis on configuration space {II}. {B}ogoliubov
  functional and equilibrium states.
\newblock In preparation, 2005.

\bibitem[KKK04]{KoKuKva}
{Yu}.~G. Kondratiev, T.~Kuna, and O.~Kutoviy.
\newblock On relations between a priori bounds for measures on configuration
  spaces.
\newblock {\em Infin.~Dimens.~Anal.~Quantum Probab.~Relat.~Top.},
  7(2):195--213, 2004.

\bibitem[KKO02]{KoKuOl00}
{Yu}.~G. Kondratiev, T.~Kuna, and M.~J. Oliveira.
\newblock Analytic aspects of {P}oissonian white noise analysis.
\newblock {\em Methods Funct.~Anal.~Topology}, 8(4):15--48, 2002.

\bibitem[KKO04]{KoKuOl00b}
{Yu}.~G. Kondratiev, T.~Kuna, and M.~J. Oliveira.
\newblock On the relations between {P}oissonian white noise analysis and
  harmonic analysis on configuration spaces.
\newblock {\em J. Funct. Anal.}, 213(1):1--30, 2004.

\bibitem[Kno64]{Kn64}
K.~Knopp.
\newblock {\em Theorie und {A}nwendung der {U}nendlichen {R}eihen}.
\newblock Springer Verlag, Berlin, Heidelberg, and New York, 5th edition, 1964.

\bibitem[KRR04]{KRR02}
{Yu}.~G. Kondratiev, A.~L. Rebenko, and M.~R{\"o}ckner.
\newblock On diffusion dynamics for continuous systems with singular
  superstable interaction.
\newblock {\em J.~Math.~Phys.}, 45(5):1826--1848, 2004.

\bibitem[Kun99]{K00}
T.~Kuna.
\newblock {\em Studies in Configuration Space Analysis and Applications}.
\newblock PhD thesis, Bonner Mathematische Schriften Nr.~324, University of
  Bonn, 1999.

\bibitem[Kun05]{Ku05}
T.~Kuna.
\newblock Bochner's theorem for {B}ogoliubov functionals.
\newblock In preparation, 2005.

\bibitem[Mec67]{Me67}
J.~Mecke.
\newblock Station\"are zuf\"allige {M}a\ss e auf lokalkompakten {A}belschen
  {G}ruppen.
\newblock {\em Z. Wahrsch. verw. Gebiete}, 9:36--58, 1967.

\bibitem[Naz85]{Naz85}
G.~I. Nazin.
\newblock Method of the generating functional.
\newblock {\em J.~Sov.~Math.}, 31:2859--2886, 1985.

\bibitem[NZ79]{NZ79}
X.~X. Nguyen and H.~Zessin.
\newblock Integral and differential characterizations of the {G}ibbs process.
\newblock {\em Math.~Nachr.}, 88:105--115, 1979.

\bibitem[Osa96]{O96}
H.~Osada.
\newblock Dirichlet form approach to infinite-dimensional {W}iener processes
  with singular integractions.
\newblock {\em Comm. Math. Phys.}, 176(1):117--131, 1996.

\bibitem[Par67]{P67}
K.~R. Parthasarathy.
\newblock {\em Probability Measures on Metric Spaces}.
\newblock Probability and Mathematical Statistics. Academic Press, New York and
  London, 1967.

\bibitem[PZ99]{PZ99}
E.~Pechersky and {Yu}. Zhukov.
\newblock Uniqueness of {G}ibbs state for nonideal gas in {$R^d$}: The case of
  pair potentials.
\newblock {\em J.~Statist.~Phys.}, 97:145--172, 1999.

\bibitem[Rue63]{Ru63}
D.~Ruelle.
\newblock Correlation functions of classical gases.
\newblock {\em Ann. Phys.}, 25:109--120, 1963.

\bibitem[Rue69]{Ru69}
D.~Ruelle.
\newblock {\em Statistical Mechanics. Rigorous Results}.
\newblock Benjamin, New York and Amsterdam, 1969.

\bibitem[Rue70]{Ru70}
D.~Ruelle.
\newblock Superstable interactions in classical statistical mechanics.
\newblock {\em Comm. Math. Phys.}, 18:127--159, 1970.

\bibitem[Sch71]{Sch71}
H.~H. Schaefer.
\newblock {\em Topological Vector Spaces}.
\newblock Springer Verlag, Berlin, Heidelberg, and New York, 1971.

\bibitem[Str59]{S59}
E.~A. Streltsova.
\newblock Non-stationary processes in electrolyte theory.
\newblock {\em Ukrainian Math.~J.}, 11:83--92, 1959.

\bibitem[Tre67]{Trv67}
F.~Treves.
\newblock {\em Topological Vector Spaces, Distributions and Kernels}.
\newblock Academic Press, New York and London, 1967.

\bibitem[Yos96]{Y96}
M.~W. Yoshida.
\newblock Construction of infinite-dimensional interacting diffusion processes
  through {D}irichlet forms.
\newblock {\em Probab. Theory Relat. Fields}, 106:265--297, 1996.

\end{thebibliography}

\end{document}